\documentclass{aa}

\usepackage{graphicx}
\usepackage{amsmath}
\usepackage{amssymb}
\usepackage{natbib}

\usepackage{txfonts}

\bibpunct{(}{)}{;}{a}{}{,}
\def\void#1{{}}

\def\be{\begin{enumerate}}
\def\ee{\end{enumerate}}
\def\bi{\begin{itemize}}
\def\ei{\end{itemize}}

\def\~{$\sim$}
\def\={$\simeq$}
    
\def\H0{H_{0} \/$}
\def\h0{H_{0} \/$}
\def\h-1{$h^{-1} \/$}
\begin{document}

\title{Galaxy clusters in the CFHTLS}

\subtitle{First matched filter candidate catalogue of the Deep
fields\thanks{Table~\ref{tab:catalogue_i} is only available in
electronic form at the CDS via anonymous ftp to cdsarc.u-strasbg.fr
(130.79.128.5) or via http://cdsweb.u-strasbg.fr/cgi-bin/qcat?J/A+A/}}

\author{L.F. Olsen\inst{1,2}
\and  C. Benoist\inst{2}
\and A. Cappi\inst{2, 3}
\and S. Maurogordato\inst{2}
\and A. Mazure\inst{4}
\and E. Slezak\inst{2}
\and C. Adami\inst{4}
\and C. Ferrari\inst{5}
\and F. Martel\inst{2}
}

\offprints{L.F. Olsen, lisbeth@dark-cosmology.dk} 

\institute{Dark Cosmology Centre, Niels Bohr Institute, 
  University of Copenhagen,
  Juliane Maries Vej 30, 2100 Copenhagen, Denmark
\and Observatoire de la C\^{o}te d'Azur, Laboratoire
  Cassiop\'ee, BP 4229, 06304 Nice Cedex 4, France
\and INAF - Osservatorio Astronomico di Bologna, Via Ranzani 1, 40127 Bologna, Italy
\and Laboratoire d'Astrophysique de Marseille, UMR 6110 CNRS-Universit\'e de Provence, BP8, 13376 Marseille Cedex 12, France
\and Institut f\"ur Astro- und Teilchen Physik, Universit\"at Innsbruck, Technikerstra\ss e 25, 6020 Innsbruck, Austria}

\date{Received -; Accepted -}

\abstract{We apply a matched-filter cluster detection algorithm to the
Canada-France-Hawaii Telescope Legacy Survey (CFHTLS) $i-$band data
for the Deep-1, Deep-2, Deep-3 and Deep-4 fields covering a total
of 4~square degrees.  To test the implemented procedure we carry out
simulations for assessing the frequency of noise peaks as well as
estimate the recovery efficiency.  We estimate that up to $z\sim0.7$
the catalogue is essentially complete for clusters of richness class
$R\gtrsim1$. The recovered redshifts are in general overestimated by
$\Delta z=0.1$ with a scatter of $\sigma_{\Delta z}\sim0.1$, except at
redshifts $z\gtrsim1$ where the estimated redshifts are systematically
underestimated.  The constructed cluster candidate catalogue contains
162 detections over an effective area of 3.112~square degrees
corresponding to a density of $\sim 52.1$ per square degree. The
median estimated redshift of the candidates is $z=0.6$. The estimated
noise frequency is $16.9\pm5.4$ detections per square degree.  From
visual inspection we identify systems that show a clear concentration
of galaxies with similar colour. These systems have a density of
$\sim20$ per square degree.  \keywords{methods: data analysis --
Surveys -- Galaxies: clusters: general -- cosmology: large-scale
structure of Universe}}

\maketitle

\section{Introduction}

Clusters of galaxies, the most massive gravitationally bound systems
in the Universe are located at the nodes of the cosmic web and are
thus good tracers of the large-scale structures
\citep[e.g.,][]{bahcall88,huchra90,collins00}. Their intimate
connection with the power spectrum of the initial density fluctuations
as well as to the cosmological parameters make them important targets
in observational cosmology \citep[e.g.  ][]{bahcall97, oukbir97,
holder01, schuecker03}. Additionally, they are still accreting matter
along filaments and even at $z=0$ a large fraction of them do present
signs of substructures in their morphology and in the galaxy velocity
distribution. Therefore, a prerequisite for addressing any
cosmological issue from a cluster sample is not only to have a large
homogeneous sample with a well-understood selection function, but also
a good understanding of cluster evolution including constraints on
their dynamical state.

The evolution of galaxy clusters has in recent years obtained a lot of
attention. This is partly due to the advent of large telescopes and
sensitive instruments which allow to carry out detailed studies of
high redshift systems. Many studies have concentrated on the colour of
the red sequence galaxies. These have shown that most of the known
systems exhibit properties similar to those of present day clusters
\citep[e.g., ][]{stanford97, stanford98, vandokkum00,lidman04}, when
taking into account the passive evolution of the stellar
populations. Most of the studied systems have been X-ray selected,
which is likely to trace the most massive systems at any
redshift. Therefore, the current studies may be biased towards old,
well-evolved systems missing the true progenitors of present day
clusters and galaxies \citep{vandokkum01a}. In order to better
understand such biases and selection effects it is important to use
different detection methods and wavelength regimes.

A number of different detection algorithms and wavelengths have been
used for identifying systems at increasingly higher redshifts.  The
largest cluster samples have been constructed using either optical
\citep[e.g.,][]{postman96,olsen99b,gladders01,depropris02,postman02,goto02,kim02,gal03,bahcall03,miller05,tago06}
or X-ray \citep[e.g.,][ and references
therein]{bohringer00,bohringer04} data. More recently, the
Sunyaev-Zeldovich effect has also been used for detecting clusters of
galaxies \citep{carlstrom02}.  Each method is based on one or at most
a few characteristics derived from known clusters and may thus
introduce biases against certain types of systems.  Therefore, it is
of utmost importance for each method to have a well-defined selection
function including a clear understanding of the built-in biases.
Furthermore, directly comparing catalogues covering the same area
extracted by different methods will yield important insight into the
different biases introduced by the different methods.

The Sloan Digital Sky Survey (SDSS) data with its large area and
multi-colour coverage, as well as complementing spectroscopic
information, has recently provided an important test-bed for a number
of optical cluster detection methods at low redshifts
\citep[e.g.,][]{kepner99,kim02,goto02}. Thorough comparisons between
the different methods have been carried out by
\cite{kim02,goto02,bahcall03}. These comparisons show that, not
surprisingly, there are always differences between the various
catalogues, part of which are caused by the different ways the
parameters such as for example richness are estimated. The SDSS data
are sufficient for detecting clusters to at most intermediate
redshifts ($z\sim0.5$). At higher redshifts only smaller data samples
have been available, such as the KPNO/Deeprange survey by
\cite{postman02} and the ESO Imaging Survey
\citep{olsen99a,olsen99b,scodeggio99}. Recently, the Red sequence
Cluster Survey by \citet[RCS,][]{gladders05} covering 100~square
degrees has been achieved thus starting to probe large volumes to high
redshift. However, a detailed comparison of the efficiency of
different methods at high-z has yet to be carried out.

For detailed comparison between different methods, wide, deep and
preferentially multi-passband homogeneous surveys are required in
order to provide the necessary data for a number of different
detection methods. The design of the Wide survey of the
Canada-France-Hawaii Telescope Legacy Survey (CFHTLS) provides a data
set that is particularly well-suited for carrying out such studies.
This survey is currently underway and is planned to cover
$\sim170$~square degrees in 5 passbands spread in 4 patches with
limiting magnitudes up to 25$^m$. It will provide the necessary ground
for building a large, well-controlled cluster candidate sample at
redshifts $z\lesssim1.3$, derived from a set of different search
techniques using the spatial as well as photometric properties in one
or more passbands.  Such catalogues will allow us to test accurately
the $z>0.5$ component of the cluster distribution. Using automatic
search techniques will allow us to build selection functions for each
of our independently extracted catalogues.  A careful comparison
between the various independently extracted cluster samples will allow
us to understand the additional difficulties in detecting clusters at
successively higher redshifts as well as improve our knowledge about
clusters at these redshifts.  Combining the catalogues from several
searches we will create a robust cluster sample well suited for both
cosmological and galaxy evolution studies.

This paper is the first in a series describing detection and
investigations of primarily the photometric properties of the
identified systems. In this first paper we describe our implementation
of the matched filter detection method which we apply to the $i-$band
data of the Deep Survey of the CFHTLS. The depth presently reached by
the Deep CFHTLS corresponds roughly to the final depth of the Wide
part of the survey. Furthermore, the small size of the Deep provides a
good test-bed for various detection algorithms for investigating the
potential of the Wide survey. The motivation for exploring the
$i-$band in this first paper is two-folded: first, the survey area is
planned to be covered by $i-$band as high priority, later followed by
the other passbands; second, most previous optical cluster searches
have been carried out in $i-$band, thus focusing on this filter
facilitates comparisons with previous surveys minimising the effect of
the variation in wavelength. In future papers we will apply the same
algorithm to data in other passbands in order to investigate how
sensitive this technique is to the choice of passband at different
redshifts. To investigate the advantages of using additional colour
information we will apply a colour search technique to the same data
and cross-compare the detections between the different methods. In
parallel we are working on detection methods based on photometric
redshift slicing to fully exploit the multi-passband data for
identifying structures, as well as on searches based on weak lensing
studies \citep[see][]{gavazzi06}. In future papers we will also
characterize the detected systems in terms of total luminosity,
concentration, shape parameters and colour properties such as the
existence and significance of the red sequence.

The present paper is structured as follows: In
Sect.~\ref{sec:galcats} we describe the galaxy catalogues used in the
present work. These catalogues are a modified version of the Terapix
CFHTLS galaxy catalogues. Sect.~\ref{sec:method} describes the
matched-filter cluster identification method which is an improved
version of that implemented for the ESO Imaging Survey
\citep{olsen99a}. This section also describes a series of simple
simulations to account for the rate of false-positives as well as the
selection function. In Sect.~\ref{sec:catalogues} we apply the
detection to the $i-$band galaxy catalogue of the four Deep fields of
the survey covering a total of $\sim4~\mathrm{square\, degrees}$ and
compare the constructed cluster catalogue to those of previous
works. In Sect.~\ref{sec:summary} we summarize our findings.

Throughout the paper we use a cosmological model with $\Omega_m=0.3$,
$\Omega_\Lambda=0.7$ and $H_0=75\mathrm{km/s/Mpc}$\footnote{We use
$h_{75}=\frac{H_0}{75\mathrm{km/s/Mpc}}$}.

\section{Galaxy catalogues}

\label{sec:galcats}

\begin{figure}
\resizebox{\columnwidth}{!}{\includegraphics{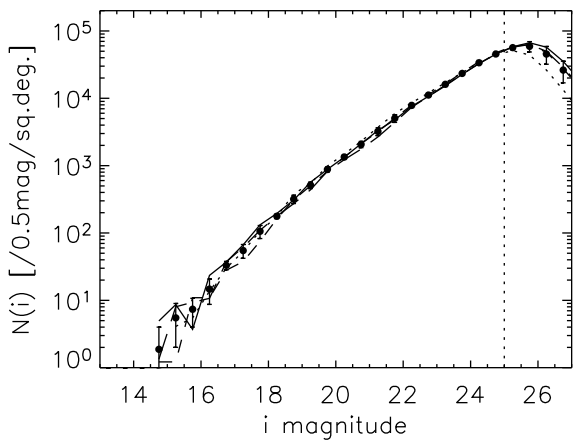}}
\caption{Average galaxy number counts (filled dots with error bars) for
the four Deep fields in the $i-$band. The
number counts for the individual fields are shown as follows: D1 --
solid line; D2 -- dotted line; D3 -- short-dashed line; and D4 --
long-dashed line. The vertical dotted lines denote our adopted
magnitude limit for the present analysis.}
\label{fig:galcounts}
\end{figure}

The galaxy samples used in this work are obtained from the catalogues
released by the Terapix team as part of the CFHTLS release T0002 in
August
2005\footnote{http://terapix.iap.fr/rubrique.php?id\_rubrique=198}.
The released catalogues are extracted with SExtractor using a
$gri$-chi-squared image \citep{szalay99} as reference and measuring
magnitudes in each individual band. Here we concentrate on the
$i-$band catalogues of the four Deep fields with a total sky coverage
of 4~square degrees. The 80\% completeness limits of the catalogues
are reached at magnitudes of $i=24.8-25.4$. Starting from the release
catalogues we apply our own star-galaxy separation based on the locus
of the objects in a half-light radius versus magnitude diagram, where
the stars are clearly separated from the galaxies at magnitudes
$i\lesssim21-22$. Furthermore, we apply a correction for Galactic
extinction based on the maps by \cite{schlegel98}.  As part of the
release, the Terapix team provides mask files for filtering out false
detections, usually caused by saturated stars or ghost images. Those
masks are produced based on the $i$-band images. We have visually
inspected the images together with the masks and, in areas where
spurious detections caused by ghost images or spikes from bright
objects were still present, defined a few additional masks.

In Fig.~\ref{fig:galcounts} we show the galaxy number counts obtained
for these post-processed catalogues for the four Deep fields. The
average counts are shown as points with error bars computed as the
standard deviation between the fields. It can be seen that the D2
field is slightly shallower than the others as was also indicated by
the 80\% completeness limiting magnitudes as given by the Terapix team
(24.8--25.4). It can be seen that the 80\% completeness limits roughly
corresponds to the peak of the counts. As a compromise we choose in
all fields to use galaxies with $i\leq25$. This limit is also
indicated in the figure.

\section{Cluster Detection}

\label{sec:method}

The cluster detection algorithm used in the present paper is based on
the matched-filter technique \citep[e.g., ][]{postman96} as it was
implemented for the ESO Imaging Survey \citep{olsen99a}. To this
version a number of improvements have been added for the present
work. Here, we summarize the main aspects of the implementation, in
particular focusing on the new implementations done for this work.

\subsection{Matched-filter cluster detection}
\label{sec:mf}

The matched-filter detection procedure is based on a maximum
likelihood analysis with the following steps:

\begin{enumerate}

\item Creation of a filter based on an assumed cluster galaxy

luminosity function and radial profile

\item Creation of likelihood maps for a series of redshifts

\item Detection of significant peaks

\item Cross-matching of peaks for different redshifts

\item Creation of likelihood curves and identification of the redshift
of maximum likelihood used for defining the cluster properties such as
redshift and richness

\end{enumerate}

The current implementation uses the \cite{postman96} matched-filter
algorithm, which filters the galaxy catalogue based on positions and
apparent magnitudes. The filter is constructed to enhance galaxy
overdensities that resemble that of a cluster with the assumed radial
profile and luminosity function (LF) embedded in a background of field
galaxies. We have adopted the Hubble radial profile characterized by a
core and a cut-off radius. The LF is a Schechter function
\citep{schechter76} characterized by the faint-end slope and Schechter
magnitude, $M^*$. The chosen values for these parameters are taken from
\cite{popesso05} and listed in Table~\ref{tab:det_parameters} after
conversion to our cosmology.  For details of the matched filter
algorithm and its variants we refer the reader to \cite{postman96,
kepner99, olsen99a} and \cite{kim02}.

\begin{table*}
\caption{Detection and filtering parameters for building the cluster
catalogues.}
\label{tab:det_parameters}
\begin{center}
\begin{tabular}{ll}
\hline\hline
Parameter & Value\\
\hline
\multicolumn{2}{c}{Filter}\\
Core radius, $r_c$ & $0.133h^{-1}_{75}\mathrm{Mpc}$\\
Cut off radius, $r_{co}$ & $1.33h^{-1}_{75}\mathrm{Mpc}$\\
Faint end slope of LF, $\alpha$ & -1.15\\
Schechter magnitude, $M^*_i$ & -22.24\\
\hline
\multicolumn{2}{c}{Likelihood maps}\\
Pixel scale & $0.5r_c$=$21.6-8.5$~arcsec\\
$z$-interval & $0.2-1.3$, $\Delta z =0.1$\\
\hline
\multicolumn{2}{c}{Peak detection}\\
Threshold & $3.5\sigma$\\
Minimum area & $\sim \pi \left(r_c(z)/2\right)^2\sim 4\mathrm{pixels}$\\
\hline
\multicolumn{2}{c}{A posteriori filtering}\\
Minimum number of shells & 2 \\
\hline
\end{tabular}
\end{center}
\end{table*}

The matched filter is applied to observed quantities (thus angular
distances and apparent magnitudes). The filter is computed in a grid
covering the galaxy catalogue, where we have chosen the pixel scale to
correspond to half the core radius. To construct the likelihood curves
the filter is tuned to a series of redshifts from $z=0.2$ to $z=1.3$
in steps of $\Delta z =0.1$ and applied to the galaxy catalogue for
each of them. The useful redshift range depends on the passband and
the depth of the data.  To derive the apparent Schechter magnitudes we
have used k-corrections derived from model spectra for an elliptical
galaxy from the Coleman library \citep{coleman80}, thus ignoring
luminosity evolution of the galaxies. The main impact of this is that
redshifts of the clusters may be biased low for the high-z part of the
sample.

As discussed in Sect.~\ref{sec:galcats}, each galaxy catalogue has an
associated mask file usually masking false detections around saturated
stars or ghost images. Since such false detections often come in
fairly dense groups, they are very efficiently detected by the cluster
finding algorithm. To avoid such false cluster detections, objects
within masked regions are discarded in the analysis.  While it is of
course important to avoid spurious cluster detections caused by false
objects within these masks, the masking of regions creates holes in
the galaxy distribution, which in turn hampers the assumption for the
matched filter algorithm of a homogeneous background. This has to be
taken into account both when estimating the background as function of
magnitude (in this case in particular the density of background
objects) and when evaluating the filter in the vicinity of the masked
areas. The lack of objects will decrease the signal compared to the
ideal case of complete coverage, thus the clusters situated in the
affected regions will have a lower probability of being detected.  As
a result of the spatial extent of the applied filter, the regions
affected by the holes are not only the masked regions but also their
immediate vicinity. Due to the larger angular extent of the filter at
low-z, the holes affect larger regions in the low-redshift shells than
in the higher redshift domain. This may in the end cause large regions
of the catalogue to be significantly incomplete.  To diminish the
effect of the holes in their vicinity we add randomly distributed
galaxies to fill the holes in the filtered galaxy catalogues before
applying the matched-filter algorithm. The density of the added
galaxies corresponds to the average density of the field.

Analogously, the signal at the edges will be decreased due to the lack
of galaxies outside the observed regions. To counteract this effect we
also add a border of randomly distributed galaxies again with a
density corresponding to the average in the field. The width of the
added frame corresponds to the extent of the largest filter radius
corresponding to the one used for $z=0.2$.  The magnitude distribution
of the added mock galaxies is the one of the catalogue itself. Of
course, this approach can never fully recover the signal of a
partially masked cluster or a cluster situated at the limit of the
survey, but it assures that the background signal is kept constant
across the field allowing a more homogeneous detection over the entire
area.

For each redshift, a likelihood map with the pixel scale corresponding
to half the core radius at this redshift is created.  The maps are
stored as FITS files to facilitate the use of standard image analysis
tools \citep[here we use SExtractor, ][]{bertin96} for detection and
characterization of the peaks. Even though the masked areas are filled,
the signal in these regions is useless, and therefore we use weight
maps to discard these regions in the peak detection.  The weight maps
are created with the same pixel grid as the likelihood maps starting
from the updated mask files for each field. To avoid splitting
clusters with substructures we do not use the deblending option of
SExtractor, even though this probably reduces the sensitivity to
clusters close to the same line of sight.

After peak detection, the significant peaks in the different maps,
corresponding to different redshifts, are cross matched using the
association program of the LDAC-tools\footnote{Leiden Data Analysis
Center,\\ ftp://ftp.strw.leidenuniv.nl/pub/ldac/software}. This tool
associates detections by their position, such that, if the detection
areas in two shells overlap, the detections are associated to each
other to create the raw likelihood functions.

Due to the slow variation of the appearance of clusters with redshift
and the luminosity width of the filter, a cluster causes a significant
peak in a number of consecutive shells. It is not likely that a
cluster absent in one shell is recovered at a more distant
redshift. Therefore, the raw likelihood functions are analysed and
split in two if it occurs that for a given redshift no detection is
found. In this way it is assured that all likelihood curves are
contiguous as is expected for real clusters.  Additionally, since the
filter size is much larger at low redshifts than at higher redshifts
it may happen that two detections are blended at low redshift but
deblended at higher redshifts. The raw likelihood curves are therefore
searched for shells where two detections are associated to the same
cluster. In such cases only the detection closest to the foreground
one is kept in the present detection and the other one used in a new
series of detections. Both of these procedures are improving the
sensitivity to clusters along the same line of sight.

The above process creates the final likelihood curve for each cluster
candidate. Again due to the persistency of real clusters we require a
detection to appear in at least two consecutive redshift bins in order
to be considered further. The properties of the cluster candidate are
determined from the redshift at which the likelihood signal is
maximized. The cluster position is taken to be the position of the
maximum signal.

Compared with the implementation for the EIS project, the main
improvements are the handling of masked regions in the galaxy
distribution as well as the diminishing of the edge effects using
randomly distributed galaxies. Also, the analysis of the likelihood
curves has been improved. The EIS implementation was using what
we here call the raw likelihood curves.

\subsection{Balancing real and false detections}

When constructing a cluster candidate catalogue, the aim is to identify
as many real systems as possible while discarding chance
alignments. As it is clear from the above description there are a
number of steps where parameters have to be selected by the user in
order to optimize this balance. When the pixel scale has been chosen,
among the remaining parameters are the detection threshold and minimum
accepted area used by SExtractor. These parameters are related to the
``raw'' peak detection in the likelihood maps. After constructing the
raw catalogues several parameters may also be useful to suppress the
noise contribution. For instance, since clusters are expected to
persist through more than one redshift shell we demand the candidates
to show up in at least two consecutive shells. Even though we use the
number of redshift bins contributing to the likelihood curve, other
parameters such as signal-to-noise or richness could also be important
discriminators.

In principle, the optimum way for addressing this point is by
using mock galaxy catalogues like the ones produced in N-body
simulations including the full hierarchy of structures. For such a
catalogue one would know the precise location of clusters as well as
their physical properties. Therefore, application of the cluster
detection to such catalogues would allow at the same time to identify
how many clusters were detected and how many additional detections,
caused by superposition effects, were found. Hence one could
immediately balance the false and real detections based on the same
simulated catalogues. However, in reality the simulation of the galaxy
distribution as well as identifying galaxy systems is a complex task,
which will be investigated in a future paper.

\begin{figure}
\resizebox{\columnwidth}{!}{\includegraphics{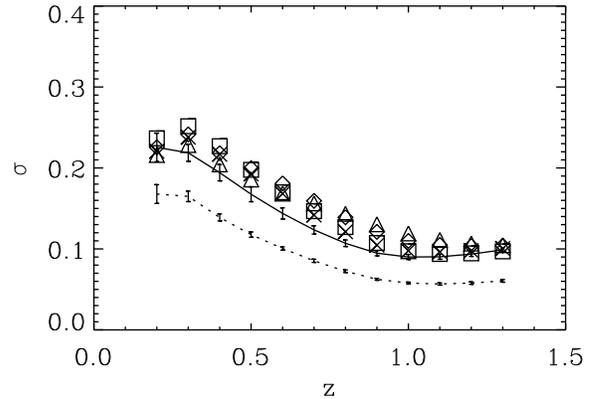}}
\caption{Variation of standard deviation of the likelihood map pixel
  value distributions as function of redshift. Each of the Deep fields
  are represented by individual symbols (diamonds for D1, triangles
  for D2, squares for D3 and crosses for D4). The lines denote the
  average over 20 mock catalogs with the error bars giving the
  standard deviation. The solid line marks the relation for the
  "correlated" background catalogues and the dotted line that of the
  "uniform" background catalogues.}
\label{fig:sigmas}
\end{figure}

Here, we adopt a simpler approach simulating separately the
distribution of the background and cluster galaxies, whereby we have
full control of the input parameters of the clusters and thus not only
the recovery rate, but also the precision of the estimated parameters
can be directly assessed.

To build backgrounds as realistic as possible means reproducing both
the magnitude distribution of galaxies and their spatial distribution
properties. In this process we encounter a coherence issue since a
fair reproduction of the clustering of galaxies (in particular at
small scales) would mean that clusters are naturally built in.
However, densities and clustering amplitudes can be reproduced in
spatially independent bins of 1 magnitude. Such a procedure allows
then to limit the luminosity coherence of a possible clump to 1 mag,
whereas a real cluster is expected to span over a much larger
magnitude range.  A set of correlated backgrounds have been
constructed based on the algorithm by \citet{soneira78}, allowing to
reproduce fairly well the slope of the two-point correlation function
as well as its amplitude as a function of magnitude. Our reference for
the number counts and correlation functions are those measured
directly from the D1 field.  Besides the correlated backgrounds we
also created for comparison spatially uniform backgrounds reproducing
only the galaxy number counts.  For each type of backgrounds 20
simulations were built.

For balancing the number of false and real detections, we applied the
matched-filter algorithm to the four Deep fields as well as to each of
the two sets of simulated catalogues. To investigate the impact
of the clustering properties on the likelihood maps, and thus on the
peak identification, we first compare the standard deviations of the
pixel value distributions. In Fig.~\ref{fig:sigmas} we show this
comparison for each redshift bin. For the simulated backgrounds we use
the mean standard deviation in each bin while for the real data each
catalogue is shown by an individual symbol. It can be seen that in
general the standard deviations of the real data are larger than those
of both of the two types of mock background catalogues. This is
expected since in the real data the presence of clusters increases the
density variations, and thus the variation in the pixel values.  While
the uniform backgrounds, as expected, show much smaller variations,
the clustered background values represent well the lower limit of the
standard deviation found in the real data.

When determining the detection threshold to be used for the peak
identification one would like to use the standard deviation of the
background pixel distribution of each likelihood map and set the
threshold in a standard ``n$\sigma$'' fashion. However, due to the
variations in clustering the background standard deviation varies as
was seen above. In order to have a common reference for the background
standard deviation, we used those derived from the average of the
20 correlated mock catalogues since these appear to represent a
clustered background well. Hence, in the following the detection
threshold scaling, $\sigma_{det}=n\sigma$, will refer to the values of
the average of the standard deviations, $\sigma$, of the 20 correlated
background mock catalogues.

\begin{figure}
\center
\resizebox{0.9\columnwidth}{!}{\includegraphics{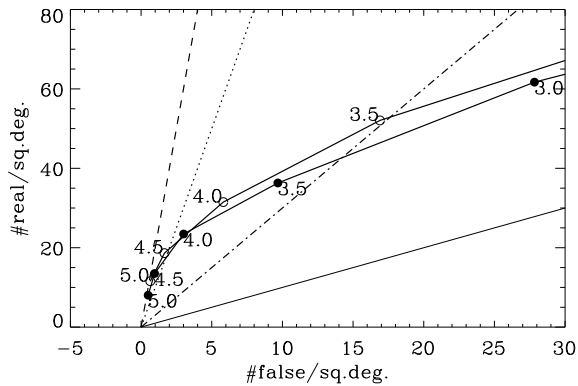}}
\caption{The number of real and false detections per square degree for
different detection thresholds and minimum area. The number of false
detections is estimated from the correlated background catalogues. The
upper curves with open symbols give the numbers for the minimum area
corresponding to $\pi (0.5r_c)^2$ and the lower curves with filled
symbols those for $\pi (1r_c)^2$. The numbers at each point mark the ``n'' in
the ``n$\sigma$'' scaling of $\sigma_{det}$. The thin solid line marks
the locus of equal number of false and real detections. The dashed,
dotted and dot-dashed lines mark a fraction of 5\% , 10\% and 33\%
false detections, respectively. }
\label{fig:false_balance}
\end{figure}

Having defined this reference it is now possible to compare the number
of detections in different cases to define the optimum detection
threshold and minimum area. The two key issues are to keep the
fraction of false-positives low, while keeping a high detection
efficiency. The first issue relates to the number of detections in the
background only simulations and the second to the number of detections
in the data catalogues. To have a fair representation of the
number of false detections we use the correlated backgrounds for these
estimates. The comparison of the number of detections in the real data
and in the correlated backgrounds is shown in
Fig.~\ref{fig:false_balance}. It shows the number of detections in
the real data versus the number of detections in the background
catalogues for different settings of the detection threshold and
minimum area. Even though the clustered background catalogues seem to
represent the background well, it is possible that the noise fraction
in this way is slightly overestimated. This (slight) overestimate is
caused by the fact that the catalogue is built in slices of one
magnitude and therefore concentrated clumps of galaxies with very
similar magnitudes may be included despite the wrong luminosity
function of the system. However, we expect this to be a minor effect,
since the correlation of position and magnitude is limited to a
relatively small magnitude range of one magnitude.

The most efficient suppression of the false detections is in the
steepest part of the curves. When it turns flatter the detections in
the catalogues become more noise contaminated.  Therefore, the most
efficient threshold judged from this relation alone is where the
curves bend off. From the figure it can be seen that the choice of
minimum area is not an efficient noise discriminator and thus to
ensure the detection of the most concentrated systems (often being
those at the highest redshifts) we choose to use the small minimum area
corresponding to $\pi (0.5r_c)^2$.  From the bending of the upper curve it can
then be seen that a threshold of $4.0\sigma$, corresponding to
$\sim$20\% false detections for a total of about $\sim$30 detections
per square degree would be the most efficient in terms of rate of
false-positives. However, other considerations such as completeness as
function of redshift have to be taken into account as well. For
investigating this we compare the resulting catalogues using both the
$4.0\sigma$ and $3.5\sigma$ detection thresholds. In the latter case
the total number of detections is $\sim$50 with $\sim30\%$ false
detections.

We have carried out two comparisons: one internal between the two
catalogues and one to an external cluster sample.  First we visually
inspected all the candidates in the $3.5\sigma$ (162 candidates) and
$4.0\sigma$ (98 candidates) catalogues and found that 122 from the
$3.5\sigma$ catalogue were graded A or B (see
Sect.~\ref{sec:catalogues} for a definition) corresponding to the most
reliable candidates. Of those candidates $\sim43\%$ are not included
in the $4.0\sigma$ catalogue. A large fraction of these missing
candidates were identified at high redshift, a main target of the
present survey. Therefore, we consider the $3.5\sigma$ threshold a
good compromise despite the increased frequency of
false-positives. This choice is supported by the results of comparing
to the XMM Large-Scale Structure Survey sample
\citep[XMM-LSS,][]{pierre06} as discussed in
Sect.~\ref{sec:xmmlss}. We find that all their $z\geq0.5$ detections
are missed if we use the $4.0\sigma$ detection threshold, but included
using $3.5\sigma$, which is our choice for the rest of this paper.

\subsection{Towards a selection function}
\label{sec:simulations}

\begin{table}
\caption{The relation between the input $\Lambda_{cl}$-richness and the
Abell richness classes. For the transformation between the counts and
richness class we use the relation found by \cite{postman96}.}
\label{tab:richness_conversion}
\begin{center}
\begin{tabular}{rrr}
\hline\hline
$\Lambda_{cl}$ & $\langle N_R \rangle$ &$R$\\
\hline
10  & 15  & $<0$\\
20  & 24  & 0\\
30  & 31  & 0\\
40  & 38  & 1\\
50  & 45  & 1\\
60  & 50  & 1\\
70  & 56  & 1\\
80  & 62  & 2\\
90  & 67  & 2\\
100 & 72  & 2\\
150 & 95  & 3\\
200 & 118 & 3\\
250 & 137 & 3\\
300 & 156 & 4\\
\hline
\end{tabular}
\end{center}
\end{table}

\begin{figure}
\resizebox{0.8\columnwidth}{!}{\includegraphics{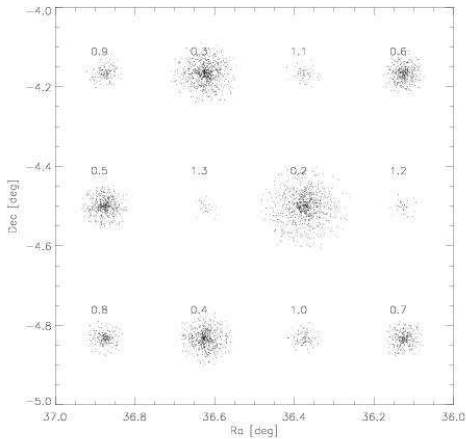}}
\resizebox{0.8\columnwidth}{!}{\includegraphics{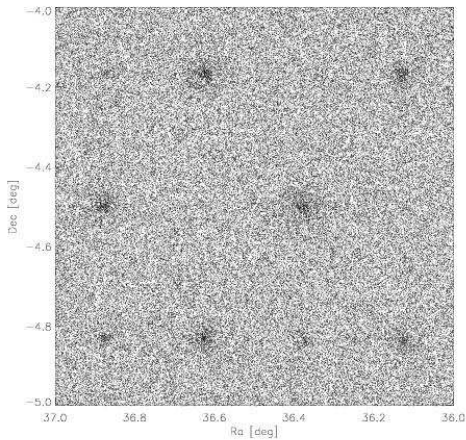}}
\resizebox{0.8\columnwidth}{!}{\includegraphics{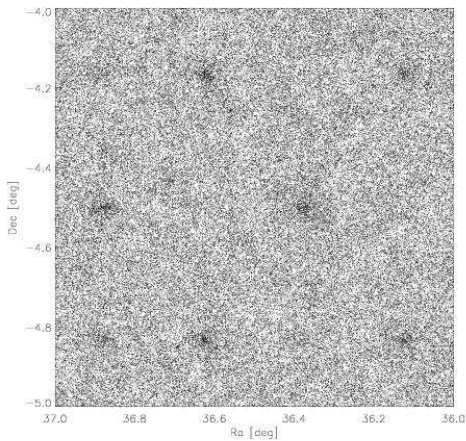}}
\caption{An example of a simulated galaxy catalogue for the clusters
only (upper panel), embedded in the uniform (middle panel) and
correlated (lower panel) backgrounds. In all cases the catalogues are
cut at the limiting magnitude of $i=25$. The clusters all have a
richness of $\Lambda_{cl}=100$ ($R\sim2$) and the redshifts indicated
in the upper panel.}
\label{fig:embed_clusters}
\end{figure}

Having determined the optimum detection parameters it is important to
investigate the recovery rate as function of redshift and richness
(selection function).  For building the selection function it is
necessary to have control on the location of clusters as well as their
basic properties. Therefore we have created simulated galaxy
catalogues based on the background catalogues described above and
clusters with the same characteristics (radial profile and LF) as the
one used for the detection filter.  A similar approach has been used
by many authors for estimating recovery efficiency
\citep[e.g. ,][]{kepner99,lobo00,goto02,kim02,postman02}.

For each background setup we create 20 mock catalogues to which we add
clusters with redshifts in the range 0.2--1.3 and
$\Lambda_{cl}$-richness from 10--300, as detailed in
Table~\ref{tab:richness_conversion}. The simulated clusters are built
to resemble the model cluster used to construct the detection
filter. In each field we add 12 clusters with different redshifts from
$z=0.2$ to $z=1.3$, but constant richness. For each richness we
construct 20 such catalogues.  Altogether a set of 280 galaxy
catalogues including clusters were produced, containing a total of
3360 galaxy clusters. Fig.~\ref{fig:embed_clusters} shows one example
of a set of 12 clusters with richness $\Lambda_{cl}=100$ ($R\sim2$)
embedded in a field background.  From the figure it can be seen that
at the highest redshifts the number of galaxies included in the
catalogue is very small, therefore the contrast of the cluster against
the background decreases significantly with redshift even for these
fairly rich systems. It is also clear that it is more difficult to
identify the clusters on the clustered background than on the uniform
one, due to the larger number of background superpositions. The
conversion between the input $\Lambda_{cl}$-richness, the Abell
richness counts, $N_R$ (number of galaxies with magnitudes in an
interval of size 2~magnitudes limited at the bright end by the third
brightest magnitude) and Abell richness class, $R$, is given in
Table~\ref{tab:richness_conversion}.

\begin{figure}
\center
\resizebox{\columnwidth}{!}{\includegraphics{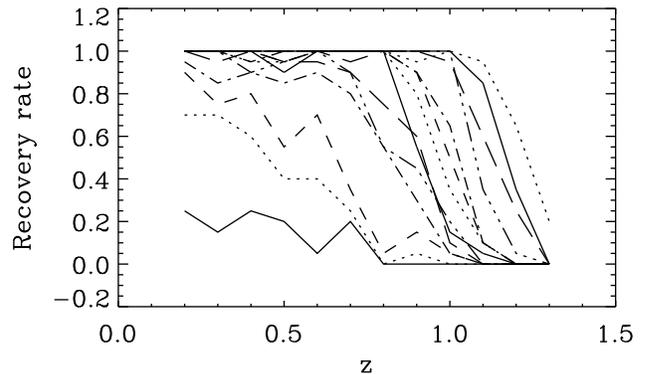}}
\caption{The detection efficiency for the correlated background
$\sigma_{det}=3.5\sigma$, area$\sim\pi (0.5r_c)^2$. The lines correspond to
$\Lambda_{cl}=10-300$ from left to right.}
\label{fig:recovery}
\end{figure}

In order to determine the recovery rate, we apply the detection
procedure to these mock galaxy catalogues and identify clusters around
the nominal cluster positions. We use a search radius of $1\farcm5$.
We reject detections that have a matching detection (in terms of
position and redshift) in the corresponding background galaxy
population. In Fig.~\ref{fig:recovery} we show the computed recovery
efficiency as function of redshift and richness. From this figure and
Table~\ref{tab:richness_conversion} it can be seen that, overall,
these settings allow clusters with Abell richness class $R\gtrsim1$ to
be detected up to redshifts $z\sim0.7$ with at least 80\% efficiency,
while at $z\gtrsim1$, systems with $R\gtrsim 2$ are detected at 80\%
completeness.

\subsection{Recovery of properties}

\begin{figure}
\resizebox{\columnwidth}{!}{\includegraphics{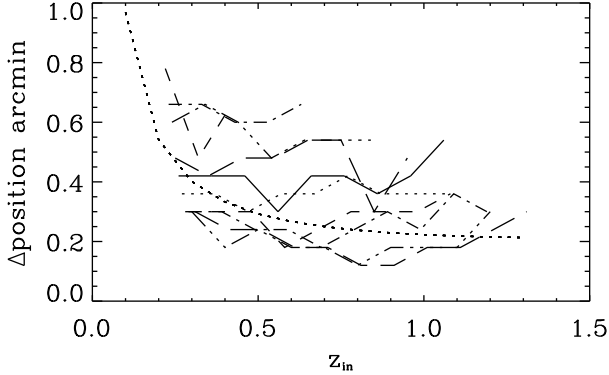}}
\caption{The average matched-filter position offset as function of
redshift and richness using a search radius of 1.5~arcmin. The
different lines correspond to different richness as in
Fig.~\ref{fig:recovery}.  The smooth dotted line gives the
corresponding angular extent of the core radius. We only include cases
with at least 50\% recovery.}
\label{fig:position_accuracy_i}
\end{figure}

In the previous section we discussed the efficiency of detecting a
cluster at the position of the simulated clusters. Another important
question for using the constructed cluster catalogue is the accuracy
of the estimated cluster properties.

The first property we investigate is the accuracy of the recovered
positions. In Fig.~\ref{fig:position_accuracy_i} we show the mean
offsets as function of redshift and richness (upper to lower curve)
and compare it with the core radius (smooth dotted line). We have chosen to
limit ourselves in redshift and richness to cases where at least 50\%
(10 cases) of the 20 input clusters are recovered. This is done in
order to make sure that we do not estimate the offsets on one special
case but obtain a reasonable statistical significance. It can be seen
that the average offsets from the nominal position is of the order or
sligthly larger than the core radius for all richnesses.  Here the
detections are recovered within $1\farcm5$ from the input center.

\begin{figure}
\center
\resizebox{0.8\columnwidth}{!}{\includegraphics{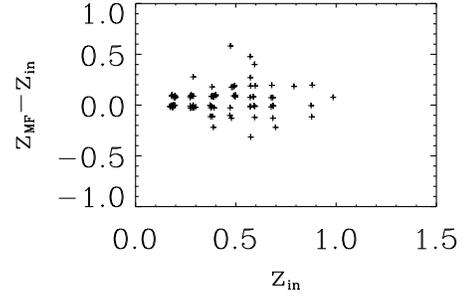}}
\resizebox{0.8\columnwidth}{!}{\includegraphics{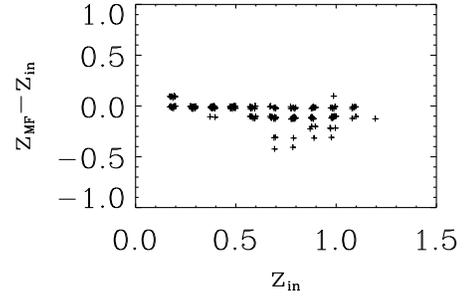}}
\resizebox{0.8\columnwidth}{!}{\includegraphics{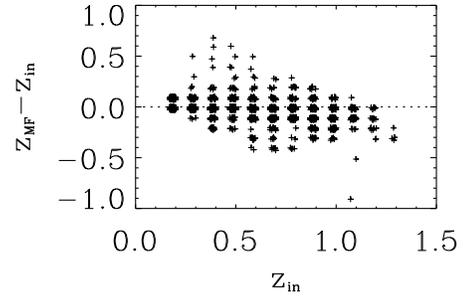}}
\caption{Redshift offsets as function of input redshifts for
richnesses of $\Lambda_{cl}=30$ (upper panel), 150 (middle panel) and
combined for all the investigated richnesses (lower panel). Each point
is slightly offset by a random number to allow all points to be seen
despite the discrete nature of the redshift measurements.}
\label{fig:redshift_accuracy}
\end{figure}

The main properties estimated by the matched filter algorithm are the
redshift and richness. To investigate the accuracy of these estimates,
we use the same matching as above adopting a search radius of
$1\farcm5$. In Fig.~\ref{fig:redshift_accuracy} we show the redshift
offsets between the input and recovered redshifts for each of the
recovered clusters mixing all richnesses in the lower panel. The two
top panels give the offsets for richnesses of $\Lambda_{cl}=30$ and
150. It can be seen that in general the recovered redshifts are in
good agreement with the input ones, even though the scatter increases
for the poorer systems. Furthermore, in the lowest redshift bins there
is a tendency of overestimating the redshift while in the highest
redshift end the opposite effect is seen. Since the mock clusters are
constructed using no-evolution k-corrections, this effect is not the
same as mentioned in Sect.~\ref{sec:mf}, but is an additional offset
introduced by the method. As discussed by \cite{schuecker98} this may
be an effect of the adopted algorithm and other choices could possibly
perform better in this respect. However, it is worth noting that
spectroscopic confirmations of clusters at low redshift detected in
the EIS program confirm the consistency between real and estimated
redshifts showing no or a slightly lower systematic offset increasing
at the highest redshifts \citep[e.g., ][]{benoist02, olsen05,
olsen05b}. The scatter is found to be consistent with the
spectroscopic results.

\begin{figure}
\center
\resizebox{0.8\columnwidth}{!}{\includegraphics{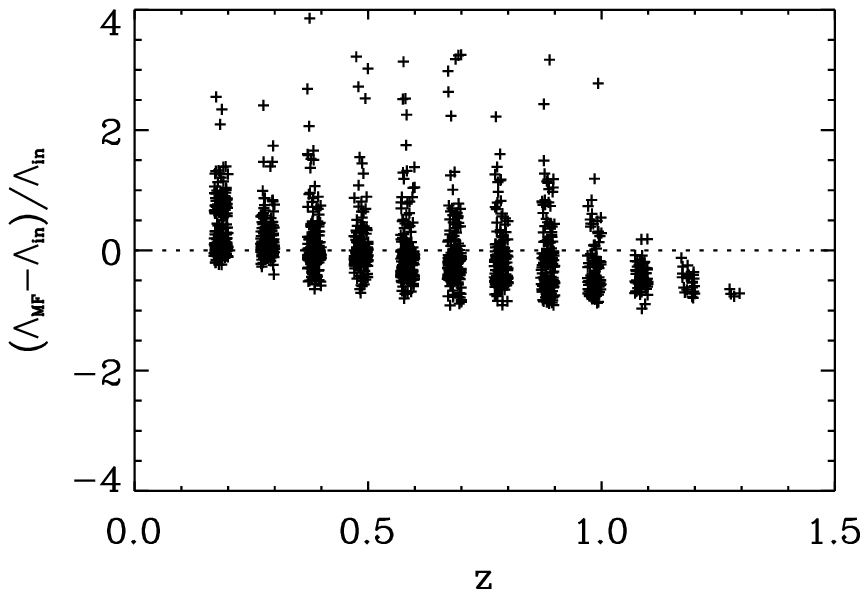}}
\resizebox{0.8\columnwidth}{!}{\includegraphics{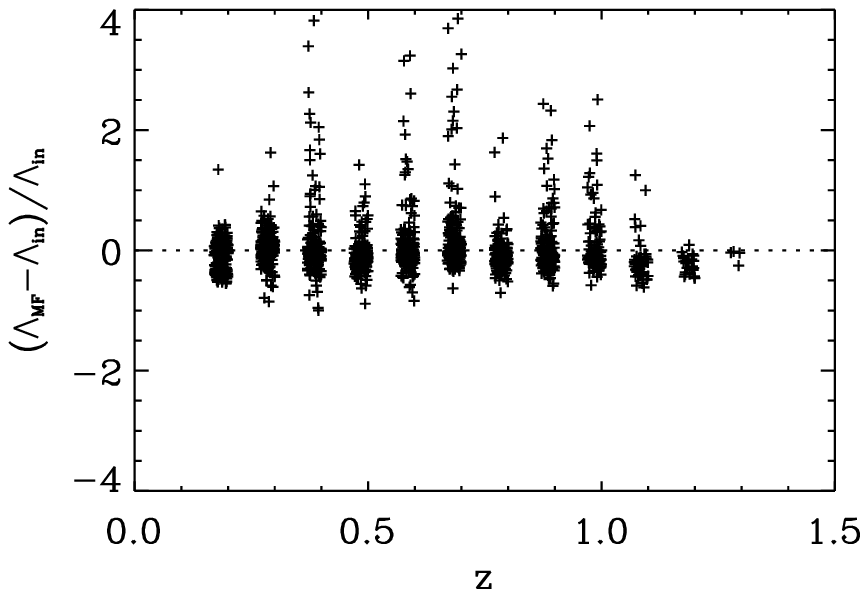}}
\caption{The relative richness offsets as function of input
redshift. The horizontal dashed line gives the zero offset line. The
upper panel shows the measured relative offsets from the input values
while the lower panels show the richness offsets after correcting for
the redshift offsets as detailed in the text.}
\label{fig:richness_accuracy}
\end{figure}

Lastly, we show the relative offsets $\left([\Lambda_{cl,
out}-\Lambda_{cl, in}]/\Lambda_{cl, in}\right)$ in the upper panel of
Fig.~\ref{fig:richness_accuracy}.  It can be seen that at low redshift
there is an overestimate on average up to 100\% and a slight
underestimate at the highest redshifts, while at intermediate
redshifts the average is in good agreement with the input richness. At
all redshifts the scatter of the recovered richness is large. The
richness estimate depends on the estimated redshift, therefore when
the redshifts are overestimated we expect the same for the richness
(an $M^*$ galaxy at higher redshift has an apparent flux that is
smaller than at lower redshift and therefore the equivalent number of
$M^*$ galaxies is larger). To investigate this effect, we corrected
the measured richness to the value corresponding to the input
redshift. This correction is made by multiplying the recovered
richness by the flux ratio between an $L^*$ galaxy at the recovered
redshift to one at the input redshift. The result of this correction
is shown in the lower panel of Fig.~\ref{fig:richness_accuracy}. It
can be seen that most of the systematic offset is removed, thus we
attribute the systematic offsets in the richness estimate to the
offset in redshift.

Even though based on simple simulations, the present results will
serve as a reference for future work investigating the effects of
clusters that do not exactly resemble the model cluster. It is an
important step for understanding the strengths and weaknesses of the
detection under well-controlled conditions before adding the
entire complexity of different cluster morphologies and superpositions
through the use of N-body simulations.

\section{Application to CFHTLS Deep Fields}
\label{sec:catalogues}

\begin{table*}
\caption{The first five entries of the cluster candidate
catalogue. The entire table is available at CDS.}
\label{tab:catalogue_i}
\begin{tabular}{lrrrrrrrr}
\hline\hline
Name & $\alpha$ (J2000) & $\delta$ (J2000) & $z_{MF}$ & $\Lambda_{cl}$ & S/N & \# Bins &  Frac. of lost area & grade \\
\hline
CFHTLS-CL-J022410-041940 & 02:24:10.3 & -04:19:40.7 & 0.9 &  79.5 &  4.32 &  3 & 0.10 & A\\
CFHTLS-CL-J022411-042511 & 02:24:11.6 & -04:25:11.1 & 1.1 & 123.0 &  4.10 &  2 & 0.08 & C\\
CFHTLS-CL-J022413-040412 & 02:24:13.4 & -04:04:12.7 & 1.1 & 120.1 &  4.01 &  2 & 0.01 & B\\
CFHTLS-CL-J022423-044044 & 02:24:23.1 & -04:40:44.3 & 0.4 &  27.2 &  4.27 &  3 & 0.03 & B\\
CFHTLS-CL-J022423-044303 & 02:24:23.3 & -04:43:03.6 & 0.5 &  29.1 &  3.90 &  4 & 0.02 & A\\
\hline
\end{tabular}
\end{table*}

\begin{figure*}
\begin{center}
\resizebox{0.45\textwidth}{!}{\includegraphics{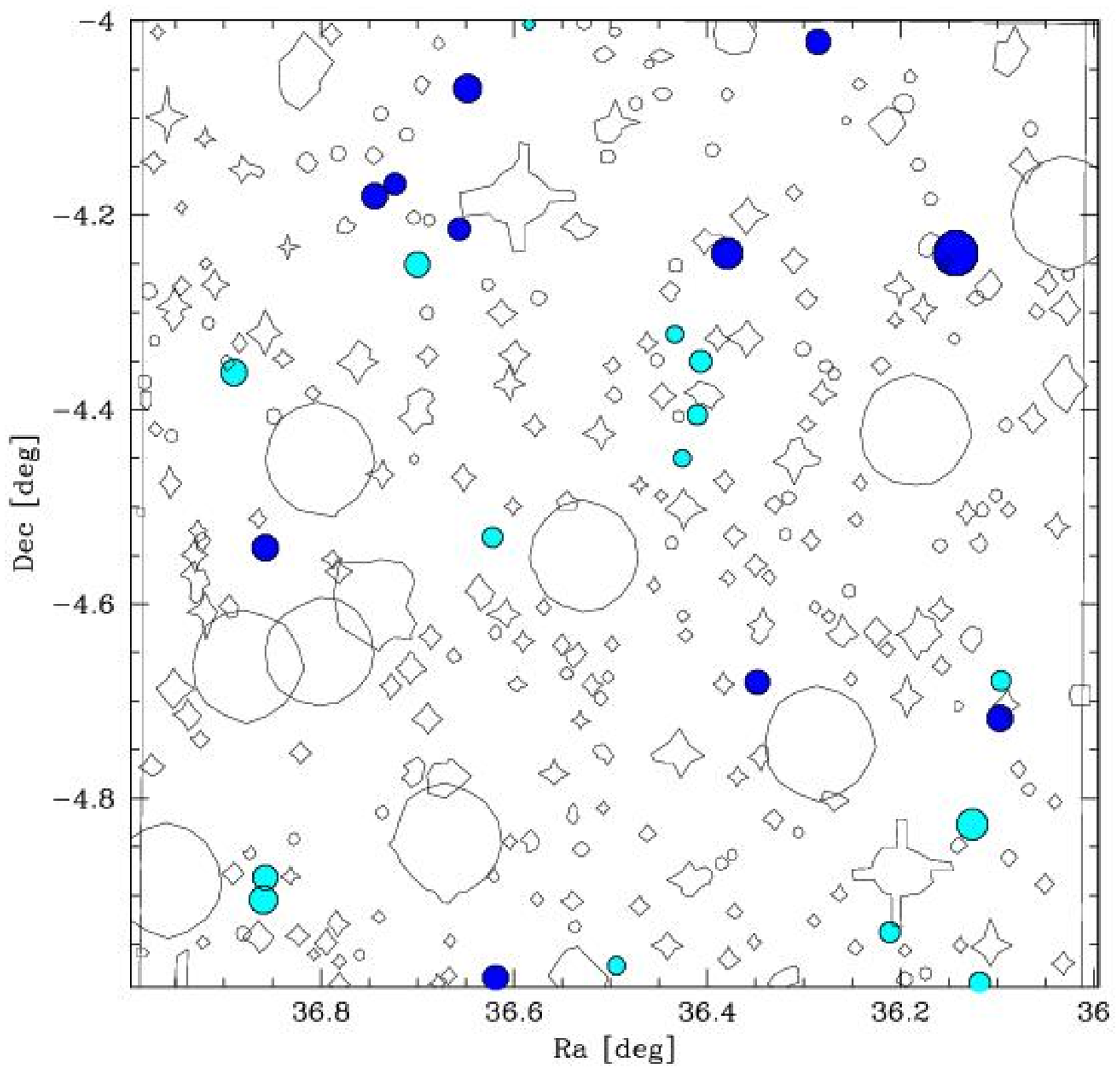}}
\resizebox{0.45\textwidth}{!}{\includegraphics{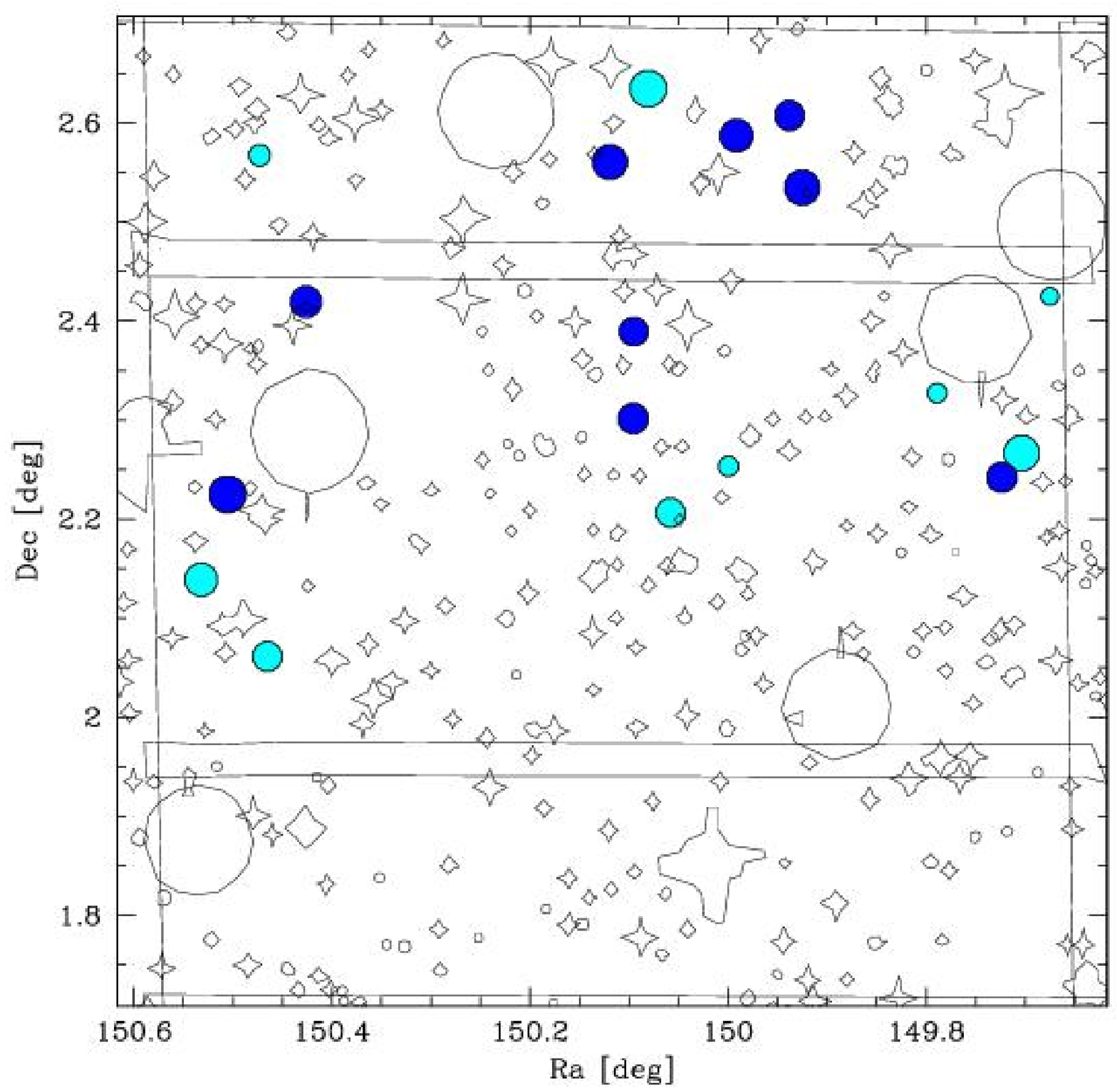}}
\resizebox{0.45\textwidth}{!}{\includegraphics{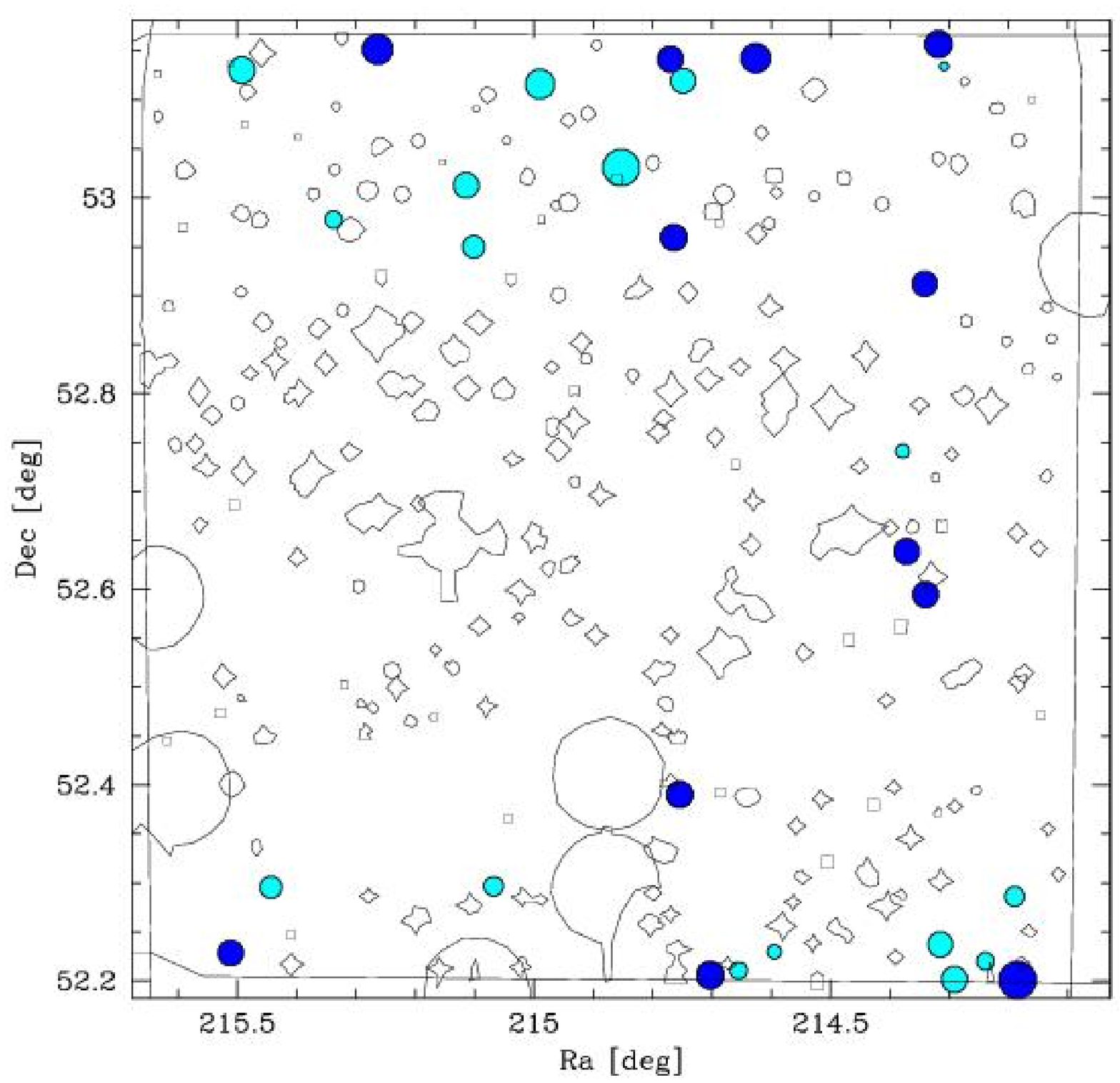}}
\resizebox{0.45\textwidth}{!}{\includegraphics{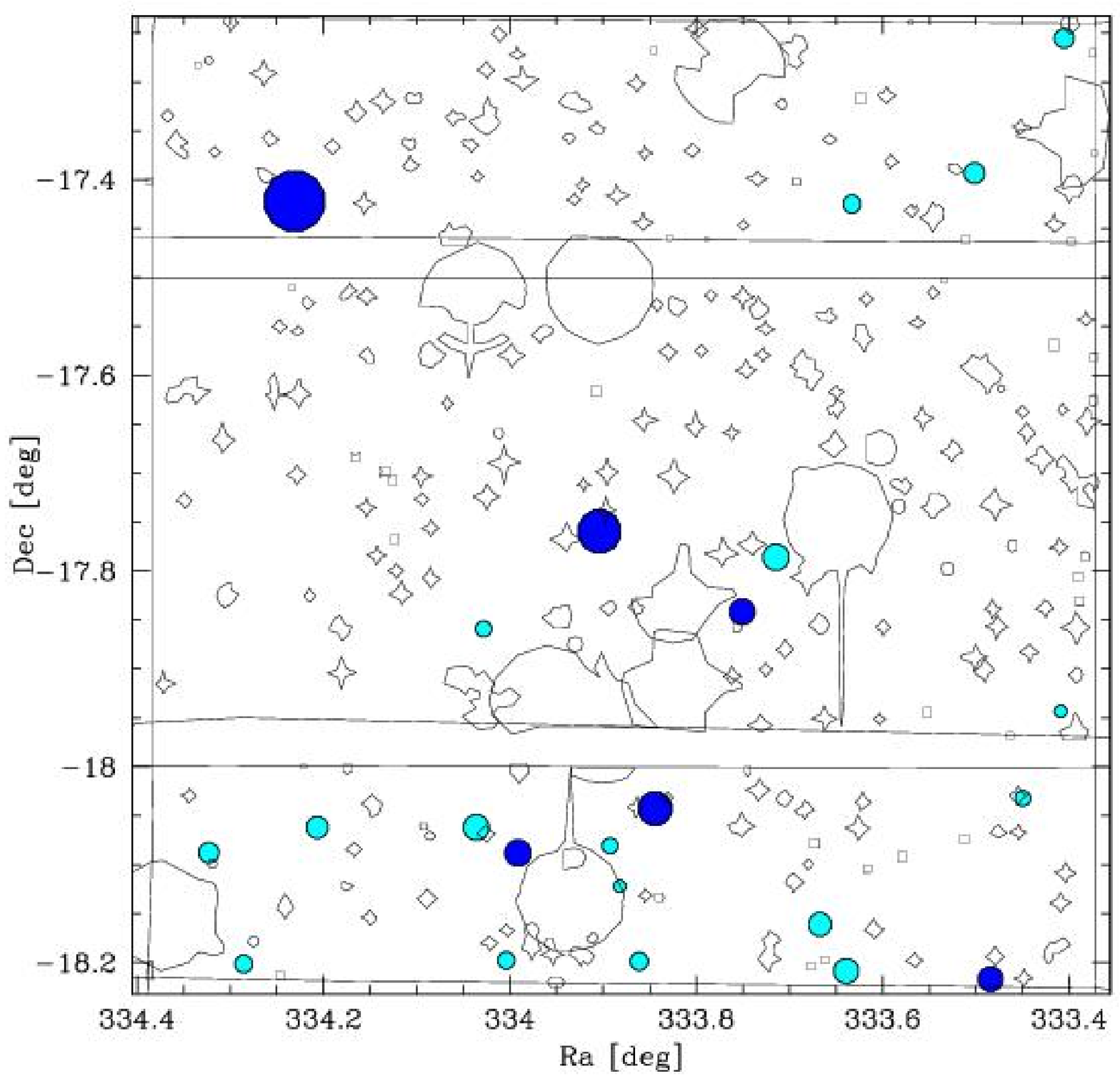}}
\end{center}
\caption{The projected distribution of clusters with $ z_{MF} \leq
  0.6$ (filled circles) detected in the four fields. The dark circles
  denote the grade A systems, while the grey ones indicate any other
  grade. The diameter of the circles increase with the richness of the
  cluster. The first row shows D1 (left) and D2 (right) and below are
  the fields D3 (left) and D4 (right). The additional polygons mark
  the position of masked regions.  }
\label{fig:spatial}
\end{figure*}

\begin{figure*}
\begin{center}
\resizebox{0.45\textwidth}{!}{\includegraphics{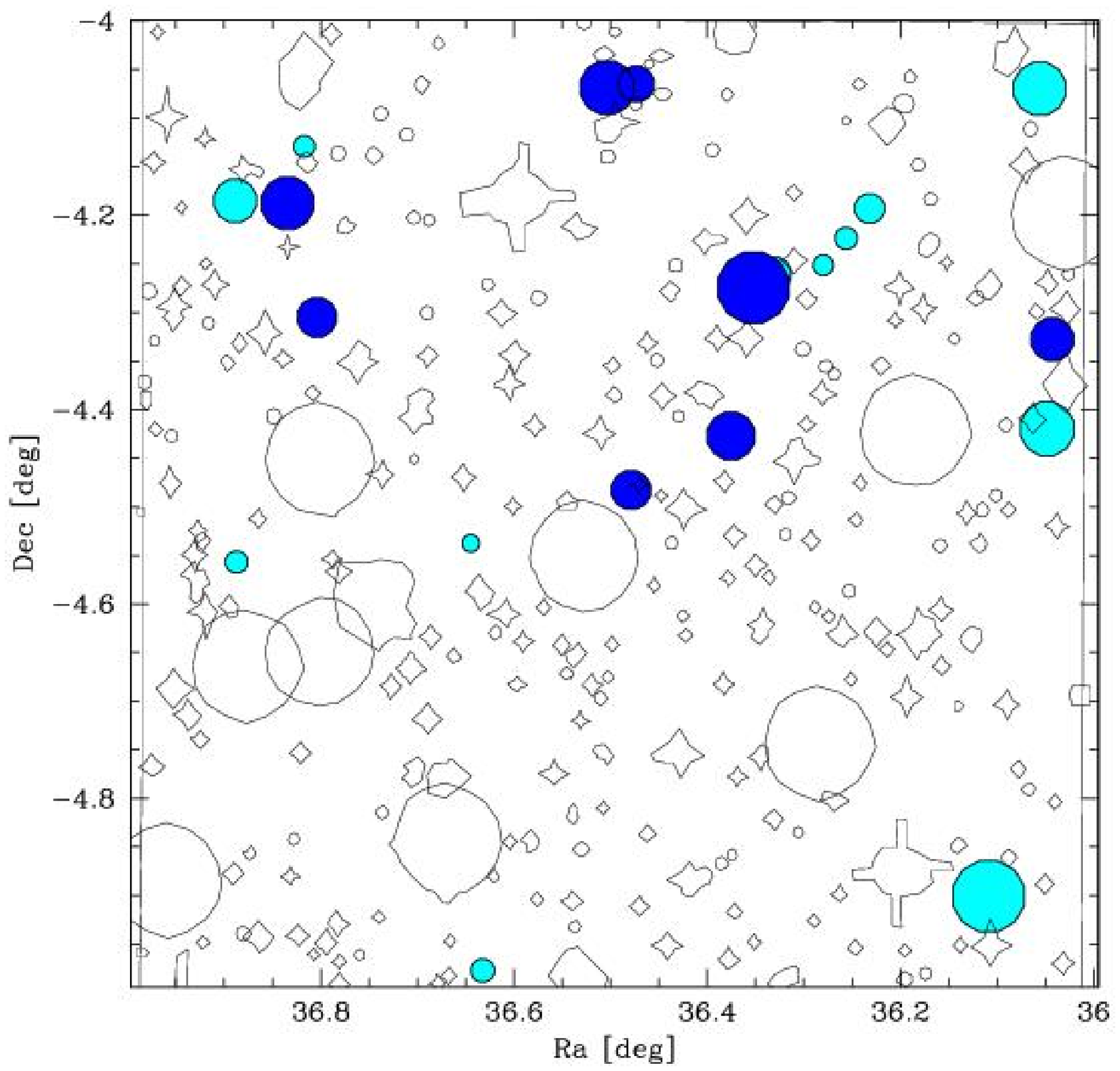}}
\resizebox{0.45\textwidth}{!}{\includegraphics{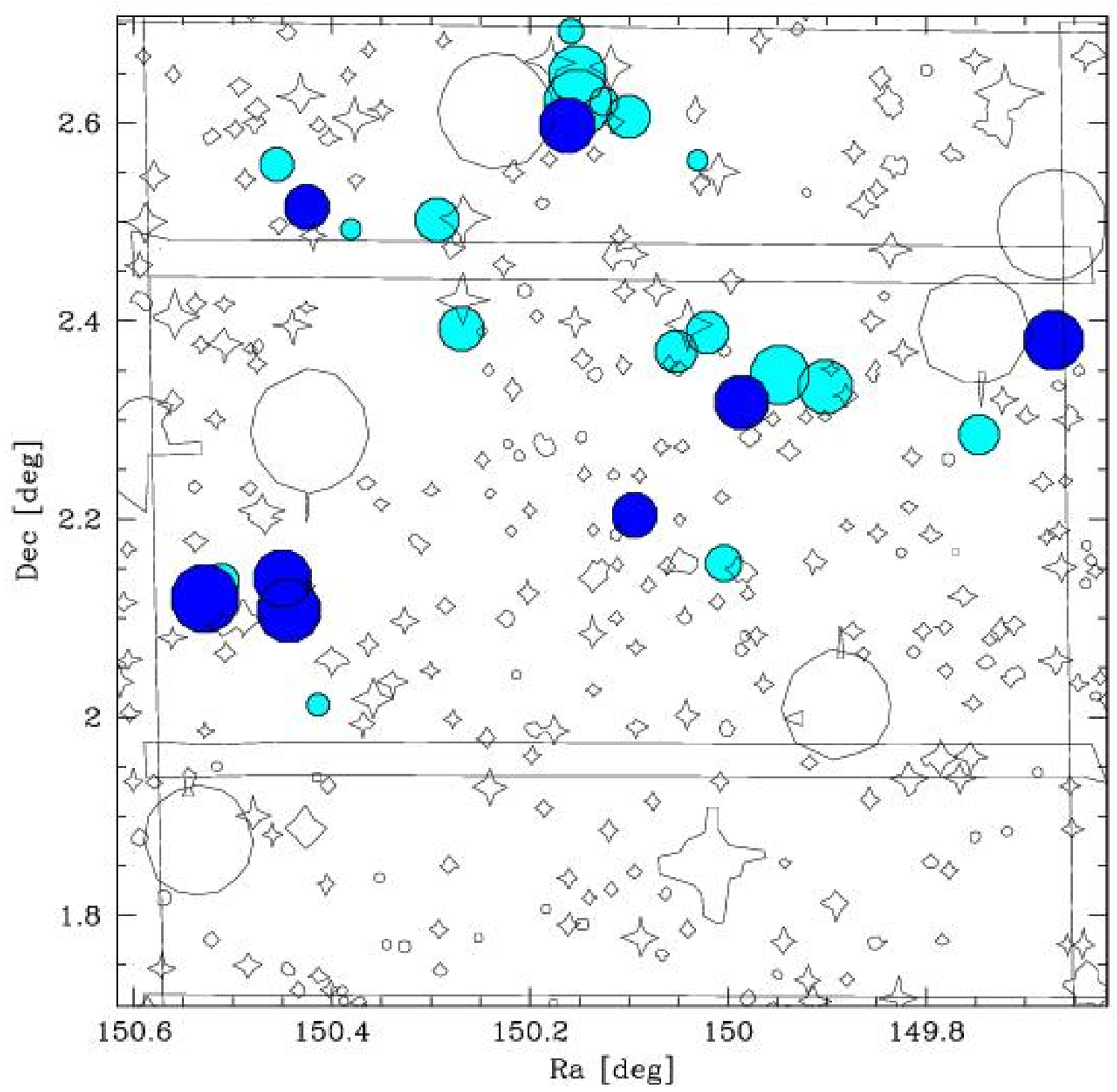}}
\resizebox{0.45\textwidth}{!}{\includegraphics{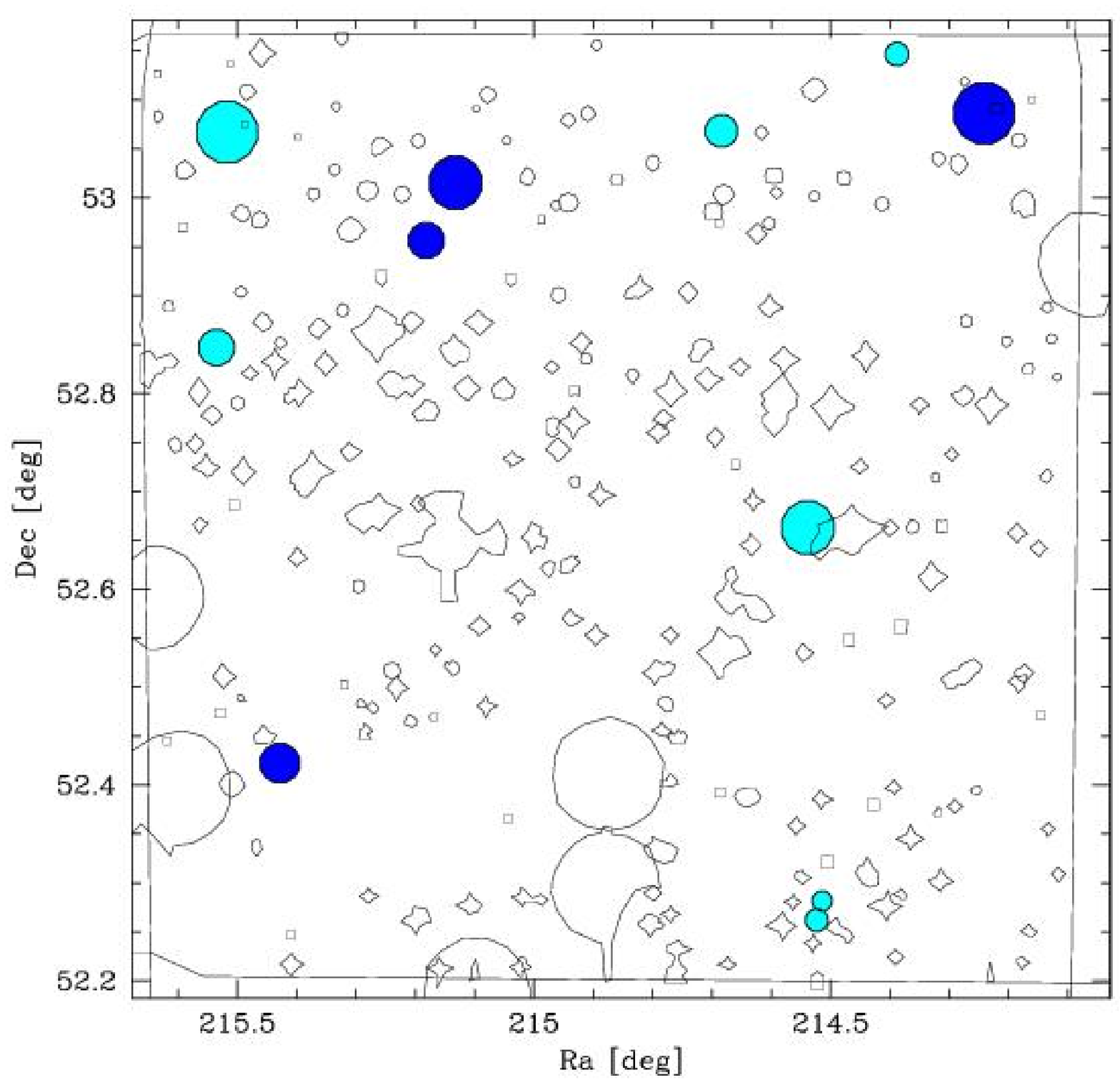}}
\resizebox{0.45\textwidth}{!}{\includegraphics{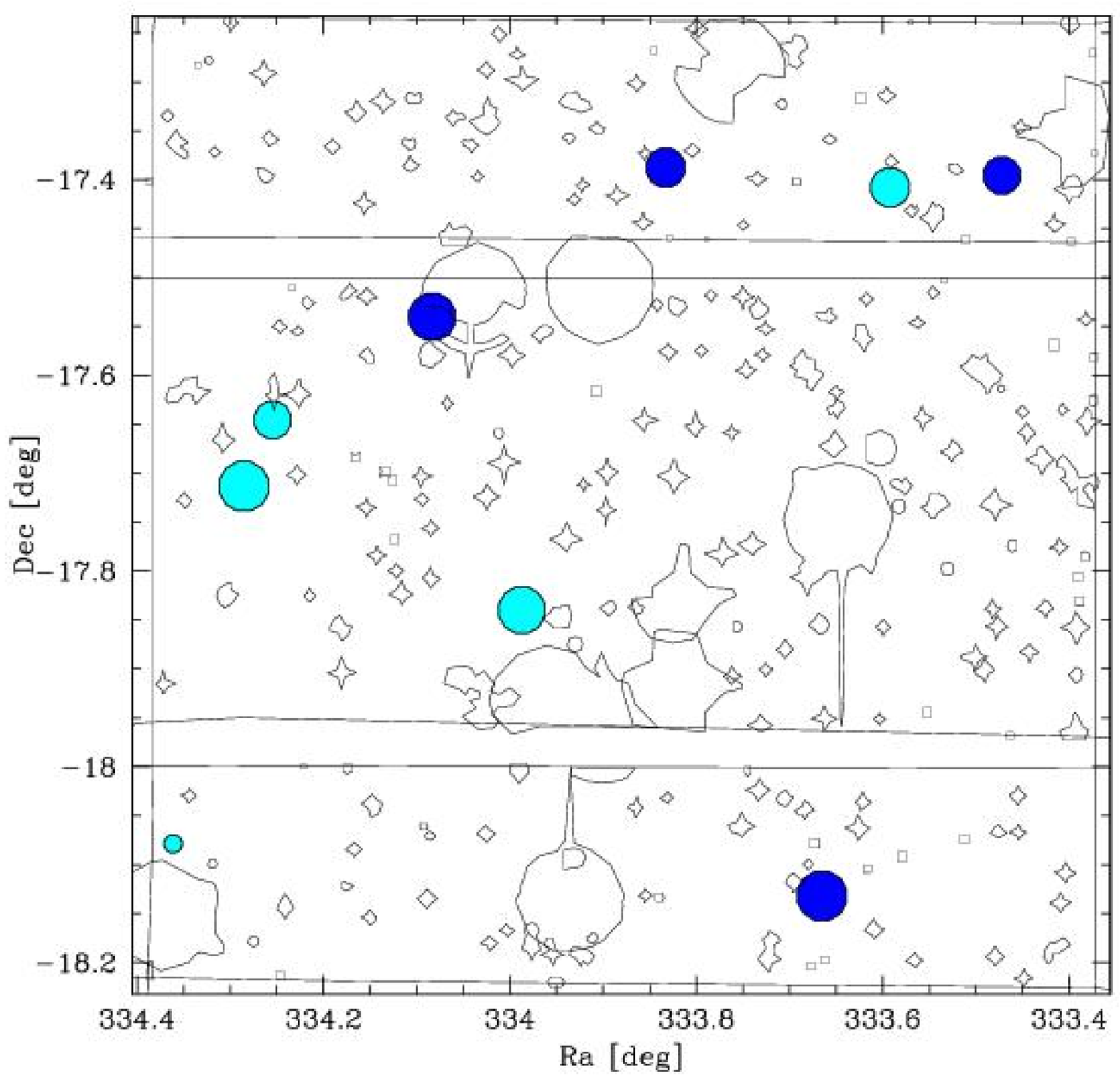}}
\end{center}
\caption{Same as Fig.~\ref{fig:spatial} but for clusters with
$z_{MF}\geq 0.7$. }
\label{fig:spatial1}
\end{figure*}

The matched filter algorithm as described above was applied to the
$i-$band catalogues of the four Deep CFHTLS fields.  The resulting
cluster candidate catalogue is presented in
Table~\ref{tab:catalogue_i} giving the first five entries. The entire
list is available at the CDS. The table lists in Col.~1 the cluster
name, in Col.~2 and 3 the right ascension and declination in J2000, in
Col.~4 the estimated redshift, in Col.~5 the $\Lambda_{cl}$ richness,
in Col.~6 the S/N of the peak value of the detection, in Col.~7 the
number of redshift bins where the candidate was detected, in Col.~8
the fraction of lost area within a distance of
$1h_{75}^{-1}\mathrm{Mpc}$ from the cluster position and in Col.~9 the
grade as defined below.  The total number of detections is 162
corresponding to 52.1 per square degree for an effective area of
$\sim3.112$~square degrees. From the density of detections in each
field we compute the average density to be $\sim52.1\pm7.8$~per square
degree, where the error indicates the scatter between the four
fields. From the simulated backgrounds we estimate to have
$\sim16.9\pm5.4$ false detections per square degree. This corresponds
to a noise fraction of $\sim$32\%. Below we compare the properties to
results of other authors.

All the detected systems were visually inspected using the related
$gri$ and $grz$ colour images. The candidates were split into four
categories denoted by the following grades: grade A systems show a
clear concentration of galaxies with similar colours; grade B systems
are characterized by an overdensity of galaxies, less concentrated
than grade A systems or without any obvious colour concentration;
grade C systems do not reveal any clear galaxy overdensity; and
finally grade D systems are systems that were detected because of lack
of masking of the galaxy catalogue or because of an artefact due to the
presence of an edge.  The relative fractions of each grade are 38.3\%
for grade A, 37.0\% for grade B, 22.8\% for grade C and 1.9\% for
grade D. From these numbers we find that the density of grade A
systems is $\sim$20 per square degree. It is worth noting that the
fraction of grade C and D systems is slightly smaller than the
estimated noise fraction possibly indicating that the correlated
backgrounds indeed slightly overestimates the true noise
frequency. However, some of the grade B systems may also be due to
superposition effects and thus contribute to the noise of the
catalogue.

In Figs.~\ref{fig:spatial} and \ref{fig:spatial1} we show the spatial
distribution of cluster candidates for each field for redshifts
$z_{MF}\leq0.6$ and $z_{MF}\geq0.7$, respectively. We show the masks
(polygons) and the detected clusters (filled circles). The size of the
circles reflects the estimated richness of the systems with larger
circles indicating richer systems. It can be seen that the number of
detections in each field varies in both redshift intervals.

\begin{figure}
\begin{center}
\resizebox{0.95\columnwidth}{!}{\includegraphics{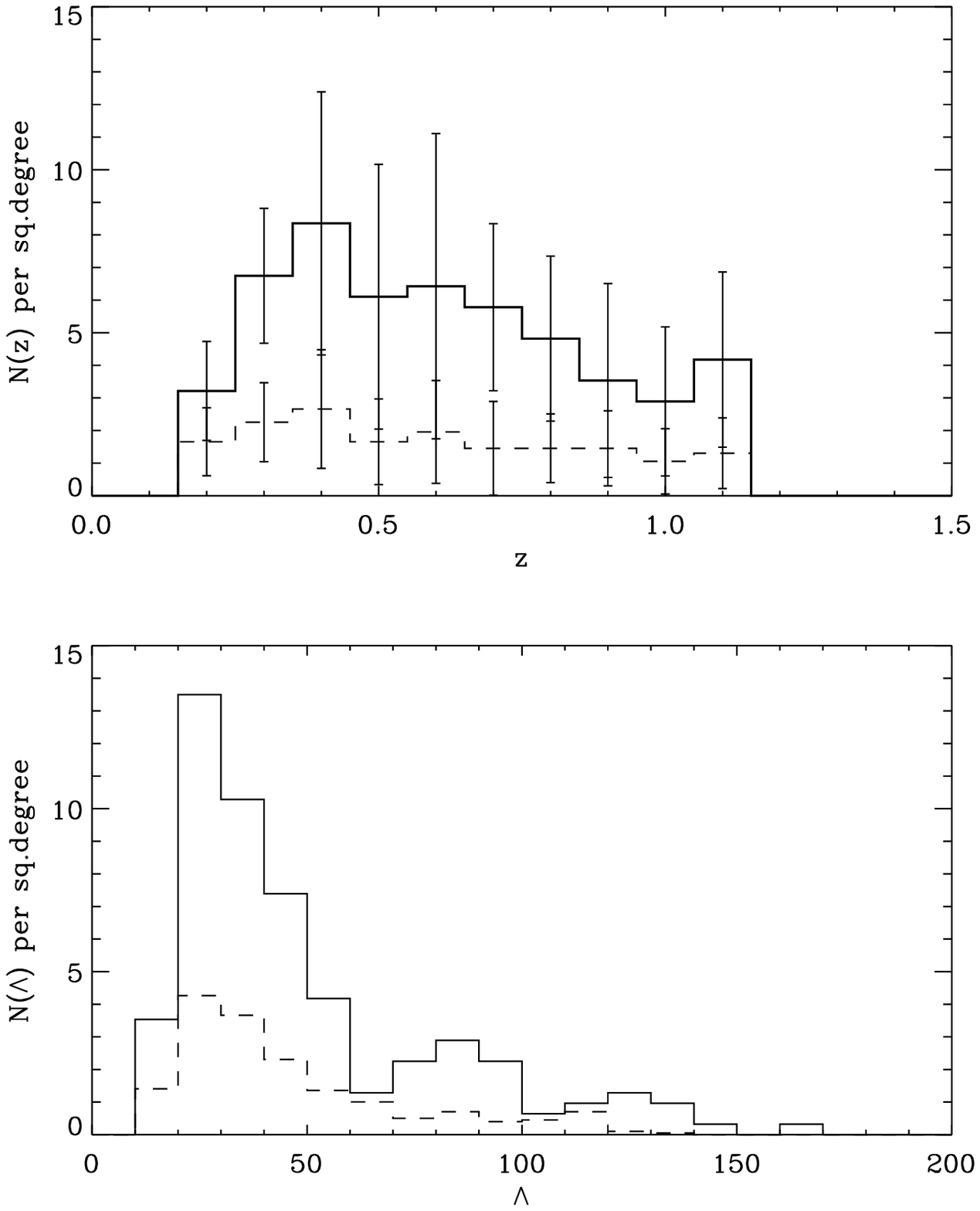}}
\end{center}
\caption{Redshift (top) and richness (lower) distributions (solid
lines) for all the candidate clusters. The distributions of false
detections (dashed lines) are estimated using the correlated
backgrounds. For the redshift distribution the error bars denote the
scatter between the fields.}
\label{fig:prop}
\end{figure}

In Fig.~\ref{fig:prop} we show the redshift and richness distributions
and compare them to the estimated contribution from false-positives.
For the redshift distribution the error bars indicate the
field-to-field variation. To investigate the expected variance of the
counts we used cluster samples extracted from N-body simulations
\citep{evrard02}. From this paper we used the results of the
$\Lambda$CDM simualtions and the deep wedge (DW) survey with a total
(simulated) sky coverage of 10$\times$10 square degrees and a maximum
redshift of 4.4. The adopted mass limits are only rough estimates
obtained from the conversion between $\Lambda_{cl}$-richness and Abell
richness class given in Table~\ref{tab:richness_conversion}. To
convert Abell richness classes to a mass limit we used the sample by
\cite{girardi98}. For each Abell richness class we computed the
average mass and used this as our rough estimate.  When extracting
cluster samples from the N-body simulations we rescaled the mass
threshold as a function of redshift to obtain samples with sizes
similar to the density in the candidate catalogue.  Extracting
randomly from these simualtions sets of 4 realisations of 1~square
degree yields a variance consistent with the field to field variation
measured from the 4 Deep fields.

\begin{figure}
\begin{center}
\resizebox{0.95\columnwidth}{!}{\includegraphics{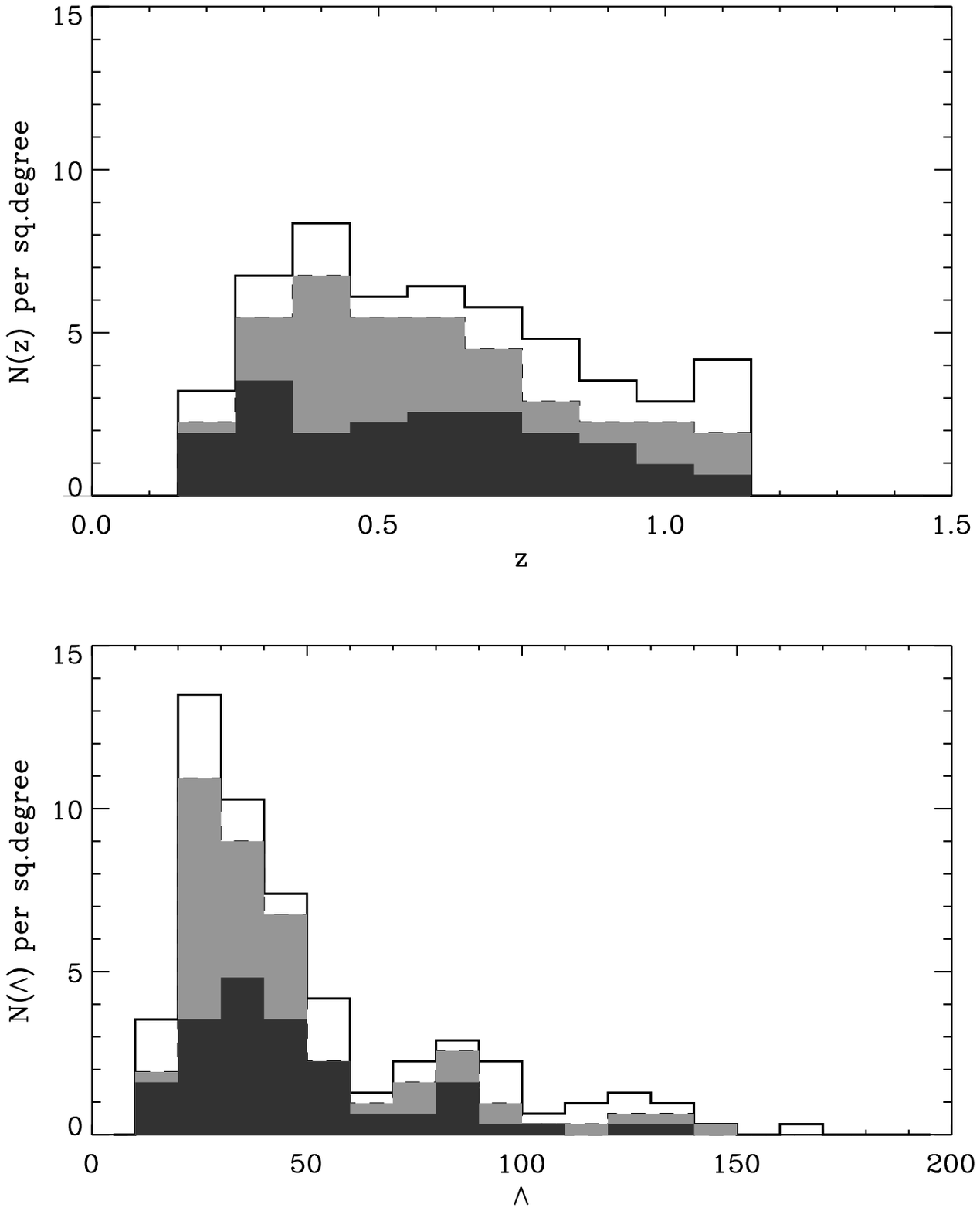}}
\end{center}
\caption{Redshift (top) and richness (lower) distributions for all the
candidate clusters (solid lines) and marking the grade A (dark grey)
and B (light grey) systems.}
\label{fig:prop_grades}
\end{figure}

Finally, Fig.~\ref{fig:prop_grades} shows the redshift and richness
distributions indicating the contribution of grade A and B systems. It
can be seen that the distributions are at any redshift and richness
dominated by grade A and B systems which are both characterized by a
clear overdensity of galaxies.

\subsection{Statistical comparison to previous surveys}
\label{sec:comp_ext}

Even though the number of high-redshift cluster searches is still
limited, a number of samples are available for comparison. First, we
compare the present catalogue with that from the EIS project
\citep{olsen99a,olsen99b,scodeggio99} since the adopted method here is
essentially the same but applied to another data set. Afterwards we
compare with two other optical catalogues: the KPNO/Deeprange
distant cluster survey using a different data set covering 16~square
degrees also in the $I-$band and using a matched-filter algorithm
\citep{postman02} and  the Red Sequence Cluster Survey
\citep[RCS; ][]{gladders05} that used both a different data set and a
different detection method, namely searching for simultaneous
overdensities in colour and space.

\begin{figure}
\resizebox{\columnwidth}{!}{\includegraphics{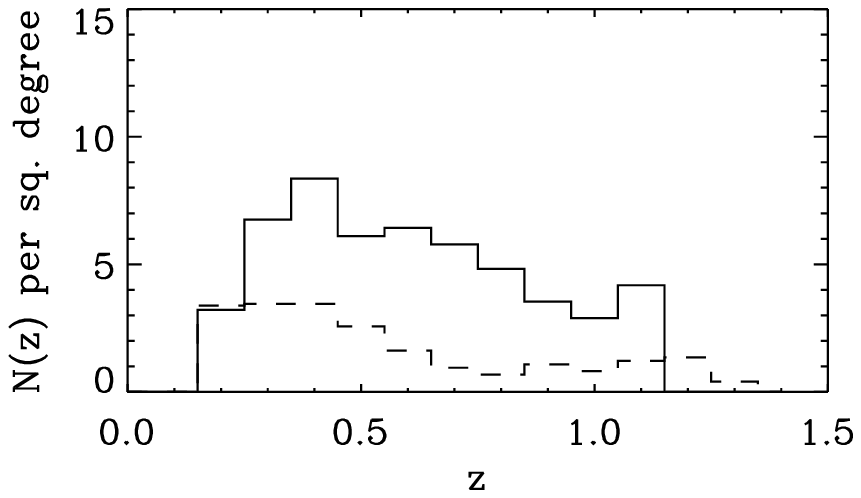}}
\caption{Comparison of redshift distributions for the
CFHTLS (solid line) and EIS (dashed line) cluster catalogues
\citep{olsen99a,olsen99b,scodeggio99}.}
\label{fig:comp_cfht_eis}
\end{figure}

\begin{figure}
\resizebox{\columnwidth}{!}{\includegraphics{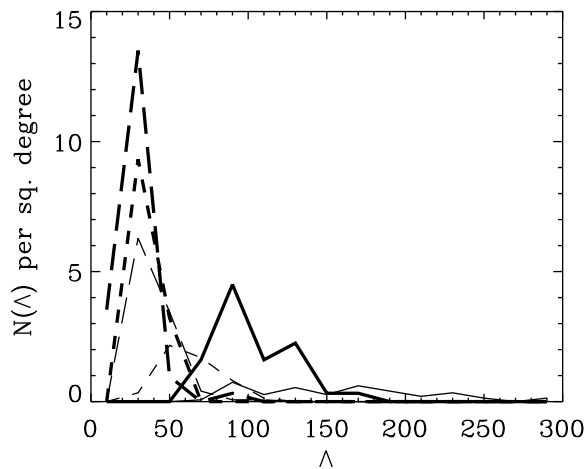}}
\caption{Comparison of richness distributions for the CFHTLS (thick
lines) and EIS (thin lines) cluster catalogues
\citep{olsen99a,olsen99b,scodeggio99}. Each line correspond to a
redshift interval as follows: long-dashed line $0.2\leq z\leq0.4$,
short-dashed line $0.5\leq z\leq0.7$ and solid $0.8\leq z\leq1.1$.}
\label{fig:comp_cfht_eis_rich}
\end{figure}

\begin{figure}
\resizebox{\columnwidth}{!}{\includegraphics{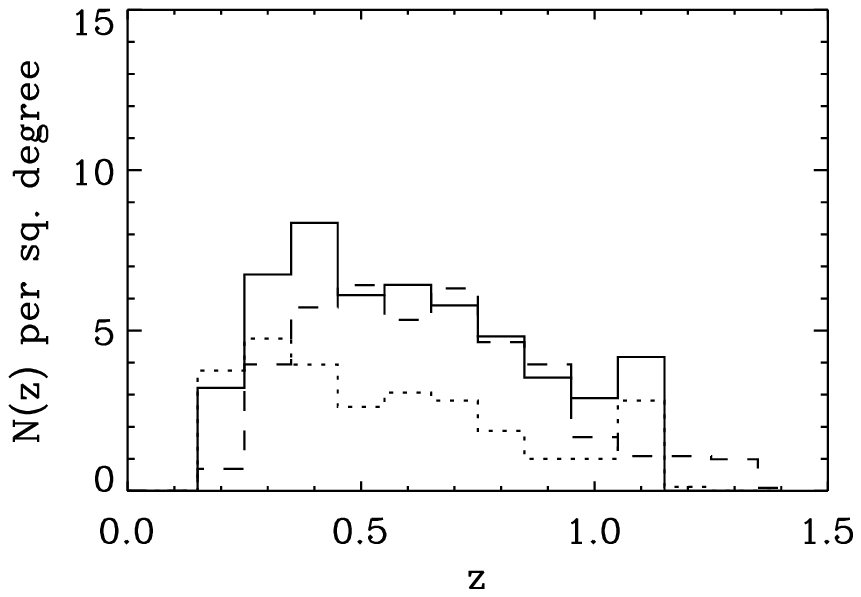}}
\caption{Comparison of redshift distributions for the CFHTLS (solid
line), KPNO/Deeprange \citep[dotted line, ][]{postman02} and RCS
\citep[dashed line, ][]{gladders05} cluster catalogues.}
\label{fig:comp_cfht_others}
\end{figure}

In Fig.~\ref{fig:comp_cfht_eis} we compare the distributions of the
properties of cluster candidates in the CFHTLS catalogue presented
here and those in the EIS cluster candidate catalogue. The difference
between the surveys is both the passband (EIS used a wide $I-$band
filter with a limiting magnitude of $I_{AB}\sim23.4$, while in
CFHTLS the more narrow Sloan $i-$band filter is used) and the depth of
the data. Already this is expected to have some impact on the obtained
cluster candidate samples, but in particular at low redshift, this is
not expected to be a large effect. The two catalogues are built with
essentially the same algorithm, however, it can be seen that the
CFHTLS includes more systems than the EIS sample at all redshift. The
large difference is likely to result from the adopted detection
thresholds, which is a major difference between the two detection
procedures. In contrast to the present project, in EIS the detection
thresholds were defined intrinsic to each field. For the richnesses
(Fig.~\ref{fig:comp_cfht_eis_rich}) we compare the detections in three
different redshift bins, since the average richness increases with
redshift (due to the selection function). Each line type corresponds
to a different redshift range as indicated in the caption. In general
the CFHTLS identifies a larger number of poorer systems than was done
in EIS which is probably a combination of the different detection
thresholds and deeper data.

In Fig.~\ref{fig:comp_cfht_others} we compare the redshift
distribution with that of the KPNO/Deeprange distant cluster survey
\citep{postman02} as well as that of the RCS \citep{gladders05}. Both
of these surveys are estimated to cover roughly the same redshift
interval as in the present survey. We find that at most redshifts the
CFHTLS distribution includes more systems than the KPNO/Deeprange
catalogue, and is comparable to the distribution of the RCS. Since the
RCS is using a different detection method also the definition of the
richness of the detected systems differs and thus a direct comparison
is not possible in this case.

For the three surveys where $\Lambda_{cl}$-richnesses are available
(CFHTLS, EIS and KPNO/Deeprange) we present the breakdown by richness
in Table~\ref{tab:rich_comp}. The richness intervals are selected
based on the values in Table~\ref{tab:richness_conversion} to
correspond roughly to Abell richness classes $R\lesssim0$, $R\sim1$,
$R\sim2$ and $R\gtrsim3$. From the table it can be seen that the
additional detections in the CFHTLS catalogue is likely due to a
larger number of relatively poor systems.

\begin{table}
\caption{For each of the three surveys the density of clusters (per
square degree) in four different richness intervals. The richness
intervals correspond roughly to Abell richness classes $R\lesssim0$,
$R\sim1$, $R\sim2$ and $R\gtrsim3$.}
\label{tab:rich_comp}
\center
\begin{tabular}{lrrr}
\hline\hline
Richness & CFHTLS & EIS & Deeprange \\
\hline
$\Lambda_{cl} < 35$ ($R\lesssim0$) & 22.5 & 3.6 & 4.2\\
$35\leq\Lambda_{cl} < 75$ ($R\sim1$) & 18.3 & 10.5 & 15.9\\
$75 \leq\Lambda_{cl} < 125$ ($R\sim2$) & 9.3 & 2.6 & 7.0 \\
$125\leq\Lambda_{cl}$ ($R\gtrsim3$)& 1.9 & 4.3 & 0.6\\
\hline
\end{tabular}
\end{table}

\subsection{Direct comparison to other samples}
\label{sec:xmmlss}

A number of clusters are already known in the investigated areas. In
order to estimate the accuracy of the estimated redshifts as well as
compare the properties of clusters identified by different methods, we
use NASA/IPAC Extragalactic Database (NED) to find clusters with known
spectrocsopic redshifts. All of these are found to be X-ray identified
systems. The main source of externally detected clusters is the
XMM-LSS \citep{valtchanov04,andreon05,willis05,pierre06} covering the
Deep-1 field. In addition one detection from the ROSAT Bright Source
Catalog Sample \citep{bauer00} is located in the Deep-4 field.

In Table~\ref{tab:overlaps} we list the previously spectroscopically
confirmed clusters found in the areas searched in the present
work. The table lists in Col.~1 a cluster id, in Col.~2 and 3 the
position in J2000, in Col.~4 the external spectroscopic redshift, in
Col.~5 the matched filter redshift when there is a match, in Col.~6 a
flag whether the cluster position falls within a mask and in Col.~7
the reference(s) of the external position of the cluster.

In the XMM-LSS catalogues of the D1 field \citep{andreon05, willis05,
pierre06} we find a total of 17 detections. When comparing to our
catalogue, we find all but 5 of these detections. Of those 5 two are
masked and the remaining three systems are optically poor.

For the 12 XMM-LSS systems where we find a counterpart and the ROSAT
cluster we compute the average redshift offsets. We find a mean offset
of $z-z_{MF}=-0.05$ with a scatter of 0.14. Among the 13 systems one
shows an offset of $\Delta z=-0.32$. For this system the colour image
shows the presence of two galaxy concentrations. The XMM-LSS position
is located between the two concentrations, while that of the present
survey corresponds to the one most distant from the XMM-LSS
detection. Thus the matching of this case is ambiguous. Discarding
this case leads to an average redshift offset of $\Delta z=-0.02$ with
a scatter of 0.12. This is consistent with the offsets estimated from
the simulations in Sect.~\ref{sec:simulations}.

All of the matched XMM-LSS systems are graded A in our catalogue.
However, the geometry of the XMM-LSS survey does not allow a complete
coverage of the D1 field. Therefore, we investigate the position of
our 9 remaining grade A systems with respect to the XMM-LSS pointings
\citep{pierre06}. Of these nine systems 3 are not covered by any
XMM-LSS pointing and 4 are found at the edge of a pointing. The last 2
systems are located in XMM-LSS pointings G01 and G02, where the G02
pointing is significantly shallower than the other pointings. The
candidate detected by our search in G01 is at $z_{MF}=0.4$ and
$\Lambda_{cl}=29$, thus not a very rich system. Also the visual
inspection of this candidate shows that the system is poor, even
though a nice concentration in colour and space is seen. Altogether,
considering only our grade A systems and the overlap with the XMM-LSS
survey, the two catalogues are in good agreement. In addition, the
present sample contains a number of systems in the same area graded B
or C for which there is no known X-ray counterparts. However, from the
visual inspection of the images we expect that more detailed
invesigations will show that a number of them corresponds to physical
systems.

\begin{table*}
\caption{Spectroscopically confirmed X-ray clusters in the four surveyed fields.}
\label{tab:overlaps}
\center
\begin{minipage}{13.5cm}
\begin{tabular}{lcccrcl}
\hline\hline
Id & Ra & Dec & $z_{spec}$ & $z_{MF}$ &Masked & Ref.\\
\hline
XLSSC029 & 36.0172 & -4.2247 & 1.05 & $-$ & Y & 1, 3\\
XLSSC044 & 36.1410 & -4.2376 & 0.26 & 0.3 & N & 1\\
XLSSJ022522.7-042648 & 36.3454 & -4.4468 & 0.46&$-$ &N & 1\\
XLSSC025 & 36.3526 & -4.6791 & 0.26 & 0.3 & N & 1\\
XLSSJ022529.6-042547 & 36.3733 & -4.4297 & 0.92 & 0.8 & N & 1\\
XLSSC041 & 36.3777 & -4.2388 & 0.14 &  0.3 & N & 1\\
XLSSC011 & 36.5403 & -4.9684 & 0.05 & 0.2 &   Y& 1\\
XLSSJ022609.9-043120 & 36.5421 & -4.5226 & 0.82 &$-$ & N &1\\
XLSSC017 & 36.6174 & -4.9967 & 0.382 & 0.5 &   Y& 4\\
XLSSC014 & 36.6411 & -4.0633 & 0.344 & 0.4 &   N & 4\\
XLSSJ022651.8-040956 & 36.7164 & -4.1661 & 0.34 & 0.3& N& 1\\
XLSSC005 & 36.7877 & -4.3002 & 1.05 & 0.9 &  N& 1, 2, 3\\
XLSSC038 & 36.8536 & -4.1920 & 0.58 & 0.9 &   N& 1\\
XLSSC013 & 36.8588 & -4.5380 & 0.31 & 0.3 &   N& 1, 4\\
XLSSC022 & 36.9178 & -4.8586 & 0.29 & $-$ & Y &1\\
XLSSJ022534.2-042535 & 36.3925 & -4.42639 & 0.92 & 0.8 & N&3\\
XLSSC005b & 36.800 & -4.23056 & 1.0 & $-$ &  N& 3\\
RBS1842 & 334.23917 & -17.42444 & 0.136 & 0.2 &  N&6\\
\hline
\end{tabular}
\\ 1. \citet{pierre06}, 2. \citet{valtchanov04}, 3. \citet{andreon05}, 4. \citet{willis05}, 5. \citet{bauer00}
\end{minipage}
\end{table*}

\section{Summary}
\label{sec:summary}

In this paper we present the first catalogue of optical cluster
candidates extracted from the CFHTLS Deep data. Using an improved
implementation of the matched-filter procedure of the EIS project
\citep{olsen99a} we construct cluster catalogues in the $i-$band for
the Deep fields of the present survey. Through simple simulations we
assess the rate of false detections as well as the recovery rate.  The
main properties of the catalogue are the following:

\begin{itemize}

\item The catalogue contains 162 clusters covering the redshift range
from $z=0.2$ to $1.1$ with a median of $z_{med}=0.6$. The density of
candidates is $52.1\pm7.8$ per square degree; among them, $\sim20$ per
square degree show a concentration of similarly coloured galaxies.

\item The estimated rate of false detections is $\sim16.9\pm5.4$ per
square degree. This density is consistent with the fraction of systems
not showing any obvious concentration in the visual inspection.

\item From simulations we find that the catalogue is complete for
systems of richness class $\geq1$ up to $z=0.7$; beyond that the
recovery rate decreases to close to zero at $z\sim1.2$.

\item We find that the estimated redshifts are in general
overestimated by $\Delta z\sim0.1$ with a scatter of $\sigma_{\Delta
z}\sim0.1$. Correcting for this redshift offset, the recovered
richnesses are in good agreement with the input.

\item The present catalogue extracted from the CFHTLS Deep fields
appears to trace poorer systems at higher redshifts than previous
matched-filter cluster catalogues.

\end{itemize}

We have compared our catalogue with that of the XMM-LSS survey
\citep{pierre06}. From this comparison we found that our grade A
systems in the region of overlap between the two surveys were detected
by XMM-LSS. The remaining X-ray detections that were not included in
our catalogue were either in masked regions or appeared optically
poor. The grade B and C systems were not included in the known (X-ray)
cluster samples.

We conclude that, the CFHTLS imaging survey provides a good basis for
constructing large samples of galaxy clusters to study the evolution
of cluster proporties as well as for other cosmological
applications. Furthermore, the ability to trace the poor cluster
population may allow for studying the processes of cluster growth and
their impact on the galaxy populations and evolution in more detail
than with previous samples. In order to gain a better understanding of
the detected systems we will use photometric redshifts to investigate
their properties. However, to fully describe their nature we have to
carry out a thorough spectroscopic follow-up.

\begin{acknowledgements}

We thank the referee, Peter Schuecker, for useful comments, which
helped improving the manuscript. During the work we have benefitted
from useful discussions with S. Arnouts, R. Gavazzi and G. Soucail.
This work is based on observations obtained with MegaPrime/MegaCam, a
joint project of CFHT and CEA/DAPNIA, at the Canada-France-Hawaii
Telescope (CFHT) which is operated by the National Research Council
(NRC) of Canada, the Institut National des Science de l'Univers of the
Centre National de la Recherche Scientifique (CNRS) of France, and the
University of Hawaii. This work is based in part on data products
produced at TERAPIX and the Canadian Astronomy Data Centre as part of
the Canada-France-Hawaii Telescope Legacy Survey, a collaborative
project of NRC and CNRS. This research has made use of the NASA/IPAC
Extragalactic Database (NED) which is operated by the Jet Propulsion
Laboratory, California Institute of Technology, under contract with
the National Aeronautics and Space Administration.  We thank the
French Programme National de Cosmologie for its support to the CFHTLS
galaxy cluster program. LFO acknowledges financial support from the
Danish Natural Sciences Research Council and the Poincar\'e fellowship
program at Observatoire de la C\^ote d'Azur.  The Dark Cosmology
Centre is funded by the Danish National Research Foundation. CB thanks
the Dark Cosmology Centre for hospitality during the final stages of
this work. AC acknowledges visitor support from the CNRS through a {\em
poste rouge} and the Observatoire de la C\^ote d'Azur for its
hospitality.

\end{acknowledgements}

\bibliographystyle{aa}

\bibliography{/home/lisbeth/tex/lisbeth_ref}

\end{document}